\documentclass[lettersize,journal]{IEEEtran}
\pdfoutput=1
\usepackage{tikz}
\usepackage{amsmath,amssymb,amsfonts}
\usepackage{algorithmic}
\usepackage{algorithm}
\usepackage{array}
\usepackage[caption=false,font=normalsize,labelfont=sf,textfont=sf]{subfig}
\usepackage{textcomp}
\usepackage{stfloats}
\usepackage{url}
\usepackage{verbatim}
\usepackage{graphicx}
\usepackage{cite}
\usepackage{float}
\usepackage{cuted}
\usepackage{flushend}
\usepackage{booktabs}
\usepackage{xcolor}
\usepackage{acronym}
\graphicspath{{fig/}}
\usepackage{mathtools}
\usepackage{amsthm}

\usepackage{ragged2e}

\newacro{ACDD}{Alamouti with cyclic delay diversity}
\newacro{URLLC}{ultra-reliable low-latency communications}
\newacro{3GPP}{third generation partnership project}
\newacro{PHY}{physical layer}
\newacro{MIMO}{multiple-input multiple-output}
\newacro{SIMO}{single-input multiple-output}
\newacro{MISO}{multiple-input single-output}
\newacro{SISO}{single-input single-output}
\newacro{MRC}{maximum-ratio combining}
\newacro{SNR}{signal-to-noise ratio}
\newacro{CP}{cyclic prefix}
\newacro{CDD}{cyclic delay diversity}
\newacro{FSC}{frequency-selective channel}
\newacro{STC}{space-time coding}
\newacro{FFT}{fast Fourier transform}
\newacro{LMMSE}{linear minimum mean-squared error}
\newacro{FER}{frame error rate}
\newacro{OFDM}{orthogonal frequency division multiplexing}
\newacro{OCDM}{orthogonal chirp division multiplexing}
\newacro{FSC}{frequency-selective channel}
\newacro{CSI}{channel state information}
\newacro{LMMSE-PIC}{linear minimum mean squared error with parallel interference cancellation}
\newacro{PFE}{perfect-feedback equalizer}
\newacro{FD}{full-duplex}
\newacro{PDP}{power delay profile}
\newacro{PDF}{probability density function}
\newacro{DFT}{discrete Fourier transform}
\newacro{SDFT}{sparse DFT}
\newacro{ICI}{inter-carrier interference}
\newacro{OTFS}{orthogonal time frequency space}
\newacro{AWGN}{additive white Gaussian noise}
\newacro{SWH}{sparse Walsh-Hadamard}
\newacro{LLR}{log-likelihood ratio}
\newacro{PMF}{probability mass function}
\newacro{CRC}{cyclic redundancy check}
\newacro{PAM}{pulse amplitude modulation}
\newacro{QAM}{quadrature amplitude modulation}
\newacro{FWHT}{fast Walsh-Hadamard transform}
\newacro{MAP}{maximum a-posteriori}
\newacro{SC}{single-carrier}
\newacro{ISI}{inter-symbol interference}
\newacro{ZP}{zero-padding}
\newacro{BCJR}{Bahl, Cocke, Jelinek, and Raviv}
\newacro{WHT}{Walsh-Hadamard transform}
\newacro{APP}{a-posteriori probability}
\newacro{SILE-EPIC}{self-iterated linear equalizer with expectation propagation}
\newacro{EP}{expectation propagation}
\newacro{i.i.d.}{independent and identically distributed}
\newacro{CWCU}{component wise conditionally unbiased}
\newacro{MSE}{mean squared error}
\newacro{EXIT}{extrinsic information transfer}
\newacro{MI}{mutual information}
\newacro{PAPR}{peak-to-average power ratio}
\newacro{DFT-s}{discrete Fourier transform-spread}
\newacro{AMP}{approximate message passing}
\newacro{GAMP}{generalized \ac{AMP}}
\newacro{VAMP}{vector \ac{AMP}}
\newacro{RSC}{recursive systematic convolutional}
\newacro{QPSK}{quadrature phase-shift keying}
\newacro{CFAR}{constant false alarm rate}
\newacro{PD}{probability of detection}
\newacro{PFA}{probability of false alarm}
\newacro{RV}{random variable}
\newacro{CDF}{cumulative distribution function}
\newacro{HD-ZP}{half-duplex ZP}
\newacro{FD-CP}{full-duplex ZP}
\newacro{DFRC}{dual-function radar communication}
\newacro{SINR}{signal-to-interference noise ratio}
\newacro{ISAC}{integrated sensing and communication}
\newacro{SI}{self-interference}
\newacro{RSI}{residual self-interference}
\newacro{RCI}{residual clutter interference}
\newacro{ADC}{analog-to-digital converter}
\newacro{ED}{energy-detection}
\newacro{ROC}{receiver operating characteristic}
\newacro{KLD}{Kullback-Leibler divergence}

\def\BibTeX{{\rm B\kern-.05em{\sc i\kern-.025em b}\kern-.08em
		T\kern-.1667em\lower.7ex\hbox{E}\kern-.125emX}}
	
\usepackage{soul,color}

\usepackage{pgfplots}
\usetikzlibrary{shapes,arrows}
\usetikzlibrary{positioning,calc}
\usetikzlibrary{decorations.pathreplacing,calligraphy}
\usetikzlibrary{arrows.meta}
\usepackage{siunitx}

\tikzset{add/.style n args={4}{
		minimum width=3mm,
		path picture={
			\draw[black] 
			(path picture bounding box.south east) -- (path picture bounding box.north west)
			(path picture bounding box.south west) -- (path picture bounding box.north east);
			\node at ($(path picture bounding box.south)+(0,0.13)$)     {\tiny #1};
			\node at ($(path picture bounding box.west)+(0.13,0)$)      {\tiny #2};
			\node at ($(path picture bounding box.north)+(0,-0.13)$)        {\tiny #3};
			\node at ($(path picture bounding box.east)+(-0.13,0)$)     {\tiny #4};
		}
	}
}

\tikzset{add2/.style n args={4}{
		minimum width=1mm,
		path picture={
			\draw[black] 
			(path picture bounding box.south) -- (path picture bounding box.north)
			(path picture bounding box.west) -- (path picture bounding box.east);
			\node at ($(path picture bounding box.south)+(0,0.13)$)     {\tiny #1};
			\node at ($(path picture bounding box.west)+(0.13,0)$)      {\tiny #2};
			\node at ($(path picture bounding box.north)+(0,-0.13)$)        {\tiny #3};
			\node at ($(path picture bounding box.east)+(-0.13,0)$)     {\tiny #4};
		}
	}
}

\newtheorem{definition}{Definition}

\definecolor{applegreen}{rgb}{0.55, 0.71, 0.0}
\definecolor{awesome}{rgb}{1.0, 0.13, 0.32}
\definecolor{azure(colorwheel)}{rgb}{0.0, 0.5, 1.0}
\definecolor{darklavender}{rgb}{0.45, 0.31, 0.59}
\definecolor{cyan(process)}{rgb}{0.0, 0.72, 0.92}
\definecolor{brightmaroon}{rgb}{0.76, 0.13, 0.28}
\definecolor{ao(english)}{rgb}{0.0, 0.5, 0.0}
\definecolor{brightturquoise}{rgb}{0.03, 0.91, 0.87}
\definecolor{bondiblue}{rgb}{0.0, 0.58, 0.71}
\definecolor{atomictangerine}{rgb}{1.0, 0.6, 0.4}
\definecolor{classicrose}{rgb}{0.98, 0.8, 0.91}
\definecolor{copperrose}{rgb}{0.6, 0.4, 0.4}
\definecolor{my_color}{RGB}{0, 0, 0} 

\begin{document}

\title{On the Performance Analysis of Zero-Padding OFDM for Monostatic ISAC Systems
 \thanks{Roberto Bomfin is with the Engineering Division, New York University
(NYU) Abu Dhabi, 129188, UAE (email: roberto.bomfin@nyu.edu).}
\thanks{Marwa Chafii is with Engineering Division, New York University (NYU)
Abu Dhabi, 129188, UAE and NYU WIRELESS, NYU Tandon School of
Engineering, Brooklyn, 11201, NY, USA (email: marwa.chafii@nyu.edu).}}

\author{Roberto Bomfin and Marwa Chafii}



\maketitle
	
\begin{abstract}
%
This paper considers an integrated sensing and communication (ISAC) system with monostatic radar functionality using a zero-padding orthogonal frequency division multiplexing (ZP-OFDM) downlink transmission.
We focus on ISAC's sensing aspect, employing an energy-detection (ED) method.
The ZP-OFDM transmission is motivated by the fact that sensing can be performed during the silent periods of the transmitter, thereby avoiding self-interference (SI) cancellation processing of the in-band full duplex operation, which is needed for the cyclic prefix (CP)-OFDM.
\textcolor{my_color}{Additionally, we also show that ZP-OFDM can reject nearby clutter interference.}
We derive the probability of detection (PD) for the ZP and CP-OFDM systems, allowing useful performance analyses.
In particular, we show that the PD expressions lead to an upper bound for the ZP-OFDM transmission, which is useful for selecting the best ZP size for a given system configuration.
We also provide an expression that allows range comparison between ZP and CP-OFDM, where we consider a general case of imperfect SI cancellation for the CP-OFDM system.
The results show that when the ZP size is 25\% of the fast Fourier transform size, the range loss of the ZP system range is only 17\% larger than the CP transmission. 
\end{abstract}


\section{Introduction}
\IEEEPARstart{I}{n} the past decades, radar and wireless communication systems have been developed independently. 
However, there are common features shared by both technologies regarding the employment of antenna arrays, operation at higher frequency bands, hardware architecture, channel characteristics, and signal processing. 
Due to these unique convergent aspects, \ac{ISAC} has been considered by the research community and industry as one of the cornerstone technologies of the future 6G radio access network \cite{LiuJSAC,CuiIoT,LiuTCOM,ZhangISAC,chafii2023twelve,BAZZI_ris,konpal,BomfinSPAWC}. 
The prospective applications made possible by ISAC are numerous, including smart factoring, the Internet of Things, robotics, environmental monitoring, and vehicular communications.
One main challenge at the core of ISAC is the unified waveform design, whose goal is to have signals that can perform both communication and sensing functionalities.
The waveform design in a unified approach can be classified into three categories, namely, sensing-centric design (SCD), communication-centric design (CCD), and joint design (JD).
The idea of SCD is to incorporate communication over an existing sensing technology \cite{Shlezinger,Zhang_ISAC}, the CCD's goal is to add sensing on top of a system designed primarily for communications \cite{Sturm,Johnston,Sayed,Sayed2}, and the JD approach is to develop new waveforms along with beamforming techniques \cite{LiuTSP,Bazzi,BAZZI_papr}.
The adoption of each strategy is application-dependent with its own pros and cons.
Another relevant design aspect is the choice of the sensing architecture in terms of uplink and downlink signaling, monostatic, bistatic, and multistatic configurations, which have their own pros and cons. The possibilities are huge and applications dependent, which makes ISAC a growing field with enormous potential to revolutionize future wireless systems.

In this work, we consider a CCD approach by employing the communication waveform based on \ac{OFDM} also for sensing \cite{Sturm,Sayed,Sayed2}.
Since OFDM is used in modern communication systems such as the 5th generation of mobile networks (5G) and the WiFi local area networks, it is beneficial to reuse the waveform of these current communication technologies for sensing.
The disadvantage is that the CCD typically favors communication over sensing, but such analysis is not in the scope of our work.
In terms of sensing architecture, we study the case of a monostatic radar functionality at a \ac{DFRC} base station with downlink signaling \cite{LiuTSP,Bazzi}.
The DFRC base station transmits downlink data to the communication users while sensing the backscattered signal. 

One main challenge of the monostatic ISAC system described above is the in-band \ac{FD} operation \cite{Everett,Barneto,Barneto2,Zeng,Clerckx,Smida}.
Since the radar transmitter is collocated to the receiver, the transmitted signal is fed to the receiver as \ac{SI} and must be properly removed otherwise the radar backscattered signal is irreparably impaired.
Basically, the reflected signal is typically orders of magnitude smaller than the SI, meaning that the requirements for the receiver components such as low noise amplifier, automatic gain control, and analog-to-digital converter are very stringent due to the huge required dynamic range.
In practice, the SI simply saturates the receiver and impedes any useful signal processing.
Thus, for typical \ac{CP}-OFDM transmissions it is mandatory to have an \ac{SI} cancellation block at the receiver.
In-band FD operation is a known problem for years and significant progress has been achieved recently.
In \cite{Everett}, $85 \,{\rm dB}$ of passive SI suppression is achieved and supported by measurements.
In \cite{Barneto}, the authors proposed a solution where they have achieved a total $75 \,{\rm dB}$ of analog isolation shown by measurements with OFDM transmission.
Another way of dealing with suppressing SI is to design beamforming techniques that create nulls between the transmitter and receiver antennas \cite{Barneto2,Clerckx}.
Furthermore, experimental works at the mmWave frequency range have shown that spacing and blockage between the transmit and receive antennas may be a possible alternative in practical deployments \cite{BomfinJCAS,BomfinSPAWC}.
Although there has been significant progress in the FD area in the past years and there have been solutions to solve the saturation problem \cite{Everett,Barneto}, in practice there is always \ac{RSI}, which can be seen as additional noise that impairs the received signal and performance.
In this work, we consider a general case of RSI, which comes from the assumption of imperfect SI processing.

The authors in \cite{Sayed,Sayed2} have proposed a rather simple and ingenious mechanism to deal with the SI problem in the in-band FD OFDM systems in ISAC.
The idea is to use \ac{ZP}-OFDM instead of CP, where the sensing part is performed during the silent periods of the transmitter.
This scheme should have minimal impact on the communication system, while completely solving the SI problem.
For example, CP-free communications systems have been reported in the literature, where techniques such as CP-restoration can be used at the receiver to enable frequency-domain signal processing \cite{BomfinTWC,BomfinCPfree}.
\textcolor{my_color}{In \cite{Sayed,Sayed2}, the authors have focused on coherent detection for the ZP signaling approach where the backscattered channel is estimated using the \ac{LMMSE}.}
Unlike \cite{Sayed,Sayed2}, we consider in this work non-coherent processing where a simple \ac{ED} technique is employed, which is an aspect of ZP-OFDM for ISAC that has not been explored yet\footnote{\textcolor{my_color}{Although \cite{Sayed2} has an estimation method termed incoherent, it is based on the LMMSE and not ED.}}.
The advantage of the ED is that it is less complex than coherent detection, and can be used in the first detection step of a sensing functionality when a multitude of spatial streams is available at the receiver.
The shortcoming is that delay and Doppler-related information are lost with ED, preventing delay-Doppler processing for parameter estimation.
Thus, the ED method can be used as a low-complexity initial detection step at the DFRC to find targets in a given spatial stream before performing parameter estimation. 

\begin{figure}\label{fig:contributions}
	\tikzstyle{block} = [draw, fill=white, rectangle, 
minimum height=1em, text width=7.4em,text centered]
\tikzstyle{block2} = [draw, fill=white, rectangle, 
minimum height=1em, text width=5em,text centered]

\tikzstyle{multiplier} = [draw,circle,fill=my_color!20,add={}{}{}{}] {} 
\tikzstyle{sum} = [draw,circle,scale=0.7,add2={}{}{}{}] {} 
\tikzstyle{input} = [coordinate]
\tikzstyle{output} = [coordinate]
\tikzstyle{pinstyle} = [pin edge={to-,thin,black}]
\newcommand\z{3}

\def\reference{(3,1)}

\def\windup{
	\tikz[remember picture,overlay]{
		\draw (-0.8,0) -- (0.8,0);
		\draw (-0.8,-0.4)--(-0.4,-0.4) -- (0.4,0.4) --(0.8,0.4);
}}

\usetikzlibrary{matrix,decorations.pathreplacing,calc,positioning,calligraphy}

\centering
\begin{tikzpicture}[auto, node distance=0cm,>=latex',scale=0.998]
	\linespread{0.8} 
	

		\draw [] (-1.4,1) rectangle (7.4,-0.65);
			\node [] at (3,0.75) {\small 1) Theoretical PD Expressions};
		\node [block2,dashed] at (-0.25,0) (sum_gamma)  {\small sum-Gamma \ref{subsec:special_case_delta0}, \ref{subsec:numerical_results_validation}};
		\node [block,dashed] at (2.65,0) (gamma) {\small Gamma \linebreak \ref{subsec:special_case_delta0}, \ref{subsec:cp_gamma_approx}, \ref{subsec:numerical_results_validation}};
		\node [block,dashed,draw = my_color] at (5.8,0) (gaussian) {\small \textcolor{my_color}{Gaussian \linebreak \ref{subsec:zp_gaussian_approx}, \ref{subsec:cp_gaussian_approx}, \ref{subsec:numerical_results_validation}}};

		\node [block] at (0,-1.5) (upper_bound) {\small 2) Upper Bound \linebreak \ref{sec:pd_zp_upper_bound}, \ref{subsec:numerical_results_upper_bound}};
		\node [block] at (3,-1.5) (range) {\small 3) ZP/CP Range \linebreak \ref{sec:range_comparison_CP_ZP}, \ref{subsec:numerical_results_range_comparison}};
		\node [block,draw = my_color] at (6,-1.5) (clutter) {\small \textcolor{my_color}{4) Clutter Rejection \linebreak \ref{subsec:clutter_rejection}, \ref{subsec:numerical_results_clutter}}};
		
		\node[] at (5.8,-1.2) (clutter2) {};
			\node[] at (2.65,-1.2) (range2) {};
			\node[] at (2.77,-0.8) (gamma2) {};
		
	
%
	\draw[->] (sum_gamma) -- (gamma) node[yshift=0.50cm,xshift=0.35cm] {};
	\draw[->] (gamma2) -| (upper_bound) node[yshift=0.50cm,xshift=0.35cm] {};
	\draw[->] (gamma) -- (gaussian) node[yshift=0.50cm,xshift=0.35cm] {};
	\draw[->] (gamma) -- (range2) {};
	\draw[->,draw = my_color] (gaussian) -- (clutter2) {};

\end{tikzpicture}
\vspace{-1.25cm}
	\caption{\textcolor{my_color}{Diagram of contributions.}}
\end{figure}
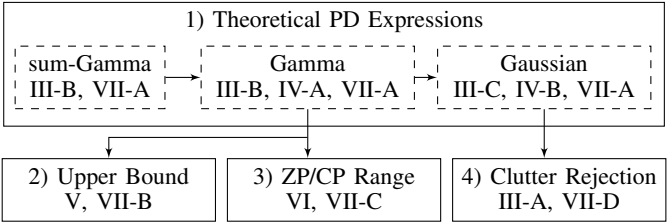	

\textcolor{my_color}{In this paper, we study ED-based monostatic sensing using ZP-OFDM in the downlink transmission.
	Our work has four main contributions and their connections are disposed in Fig.~\ref{fig:contributions}:}
\begin{enumerate}
	\item \textcolor{my_color}{The first contribution is to derive the theoretical \ac{PD} expressions for both ZP and OFDM transmissions based on sum-Gamma, Gamma, and Gaussian distributions. 
		The sum-Gamma is exact when there is no clutter, and is used to derive the approximate Gamma expression, which is also exact in some special cases.
		Moreover, we use the moments of Gamma to derive the approximate Gaussian expression, which is more accurate in the presence of clutter, and it is used to evaluate this condition.}
	
	\item \textcolor{my_color}{The second contribution focuses on studying the impact of ZP size for the scenario without clutter. 
		Basically, we show that the Gamma-based PD expression leads to an upper bound over distance for the ZP-OFDM transmission, which allows an assessment of how good a particular ZP size is in a system configuration.
		In practice, this result can serve to define the numerology of future communication systems taking into account the ED aspect for ZP transmission.}
	
	\item \textcolor{my_color}{The third contribution is to present an analytical approach that compares the detection range of the ZP and CP systems without clutter based on the Gamma distribution.
		Here we note that the ZP-OFDM transmission has the disadvantage of collecting samples only within the ZP window, while CP-OFDM has its whole transmission time to collect the backscattered signal, under the assumption of a module performing interference cancellation.
		Notably, under perfect SI cancellation, CP-OFDM certainly outperforms ZP-OFDM in terms of PD and range.
		However, there are two important questions that arise from the above observation, i) how much better is the CP-OFDM detection performance with perfect SI suppression over ZP-OFDM? and ii) how bad the SI suppression has to be for the ZP-OFDM to perform equally or better than the CP-OFDM system?
		We provide an analytical tool to answer these questions, where different levels of \ac{RSI} are considered.
		This is particularly relevant because there can be cases where the RSI is so high that the detection range of CP-OFDM becomes smaller than ZP-OFDM, even though ZP-OFDM collects data in a smaller window size.
		By proper manipulation of the PD expressions for ZP and CP transmissions, we are able to quantify how good the interference cancellation of the CP-OFDM system should be in order to outperform the ZP-OFDM system in terms of detection range.
		Ultimately, the analysis provides an assessment of whether implementing the CP-OFDM system with FD cancellation or ZP-OFDM is more advantageous.}
	
	\item \textcolor{my_color}{Lastly, we analyze the nearby clutter rejection capability of ZP-OFDM by using the PD expression based on the Gaussian approximation, assuming that the clutter is closer than the target.
		Essentially, the clutter interference has a very similar effect to the RSI due to FD, allowing the radar to neglect the samples contaminated by the nearby clutter interference. When the clutter is farther than the target, then ZP and CP-OFDM are affected similarly.}
\end{enumerate}

The remainder of this paper is organized as follows.
The system model is described in Section \ref{sec:system_model}.
In Section \ref{sec:statistics_zp}, the exact and approximate PD expressions for the ZP-OFDM system are provided.
Similarly, exact and approximate PD expressions for the CP-OFDM system are derived in Section \ref{sec:statistics_cp}.
Section~\ref{sec:pd_zp_upper_bound} presents an upper bound PD expression for the ZP-OFDM system.
In Section \ref{sec:range_comparison_CP_ZP}, an expression of the ratio between the ranges of the CP and ZP-OFDM systems is given.
The numerical results are given in Section~\ref{sec:numerical_results}.
Lastly, Section~\ref{sec:conclusion} concludes the paper.

{\it Notation}: 
a vector is described in bold letter as $\mathbf{x}$ whose $n$th element is extracted as $\mathbf{x}[n]$.
The transpose is represented by $(\cdot)^{\rm T}$. The complex conjugate of a complex number is represented by $(\cdot)^\dagger$.
The expectation and variance of the \ac{RV} $Z$ is given by $\mathbb{E}(Z)$ and $\mathbb{V}(Z) = \mathbb{E}(Z^2) - \mathbb{E}(Z)^2$, respectively.
For two \acp{RV} $Z_1$ and $Z_2$, their covariance is given by  $\mathbb{C}(Z_1,Z_2) = \mathbb{E}(Z_1Z_2) - \mathbb{E}(Z_1)\mathbb{E}(Z_2)$.  
\textcolor{my_color}{The functions $\max(a,b)$ and $\min(a,b)$ return the maximum and minimum values between the real numbers $a$ and $b$, respectively.}

\section{System Model}\label{sec:system_model}
\subsection{Transmitted Signal}\label{sec:transmitted_signal}
In this work, we consider a Gaussian distributed transmitted signal that represents an \ac{OFDM} transmission, which has either \ac{ZP} and \ac{CP}.
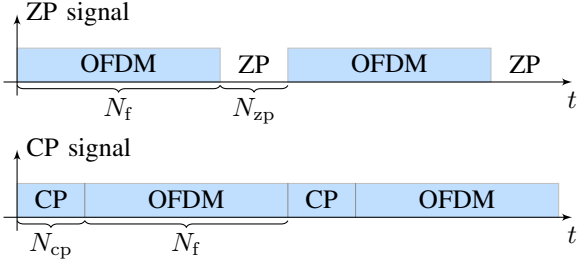
\begin{figure}[t]
	\centering
	\tikzstyle{block} = [draw, fill=white, rectangle, 
minimum height=3em, minimum width=3em]
\tikzstyle{block2} = [draw, fill=white, rectangle, 
minimum height=1em, minimum width=2.4em]

\tikzstyle{multiplier} = [draw,circle,fill=blue!20,add={}{}{}{}] {} 
\tikzstyle{sum} = [draw,circle,scale=0.7,add2={}{}{}{}] {} 
\tikzstyle{input} = [coordinate]
\tikzstyle{output} = [coordinate]
\tikzstyle{pinstyle} = [pin edge={to-,thin,black}]
\newcommand\z{3}

\def\windup{
	\tikz[remember picture,overlay]{
		\draw (-0.8,0) -- (0.8,0);
		\draw (-0.8,-0.4)--(-0.4,-0.4) -- (0.4,0.4) --(0.8,0.4);
}}


\centering
\begin{tikzpicture}[auto, node distance=0cm,>=latex',scale=0.9]
	
	\draw[->] (-0.2, 0) -- (8.2, 0) node[below] {$t$};
	\draw[->] (0, -0.2) -- (0, 1) node[right] {ZP signal};
	\draw [fill=azure(colorwheel),nearly transparent] (0,0) rectangle (3,0.5);
	\draw [fill=azure(colorwheel),nearly transparent] (4,0) rectangle (7,0.5);
	
	\node[above] at (1.5,0) {OFDM};
	\node[above] at (3.5,0) {ZP};
	\node[above] at (5.5,0) {OFDM};
	\node[above] at (7.5,0) {ZP};
	
	\draw [decorate, decoration = {brace}] (3,-0.05) --  (0,-0.05);
	\draw [decorate, decoration = {brace}] (4,-0.05) --  (3,-0.05);
	
	\node[below] at (1.5,-0.1) { $N_{\rm f}$};
	\node[below] at (3.5,-0.1) { $N_{\rm zp}$};

	\draw[->] (-0.2, 0-2) -- (8.2, 0-2) node[below] {$t$};
	\draw[->] (0, -0.2-2) -- (0, 1-2) node[right] {CP signal};
	\draw [fill=azure(colorwheel),nearly transparent] (0,-2) rectangle (1,0.5-2);
	\draw [fill=azure(colorwheel),nearly transparent] (1,-2) rectangle (4,0.5-2);
	\draw [fill=azure(colorwheel),nearly transparent] (4,-2) rectangle (5,0.5-2);
	\draw [fill=azure(colorwheel),nearly transparent] (5,-2) rectangle (8,0.5-2);
	
	\node[above] at (0.5,-2) {CP};
	\node[above] at (2.5,-2) {OFDM};
	\node[above] at (0.5+4,-2) {CP};
	\node[above] at (2.5+4,-2) {OFDM};
	
	\draw [decorate, decoration = {brace}] (1,-0.05-2) --  (0,-0.05-2);
	\node[below] at (0.5,-0.1-2) { $N_{\rm cp}$};
		
	\draw [decorate, decoration = {brace}] (4,-0.05-2) --  (1,-0.05-2);
	\node[below] at (2.5,-0.1-2) { $N_{\rm f}$};	
	
\end{tikzpicture}
	\vspace{-0.3cm}
\vspace{-0.7cm}
	\caption{ZP and CP transmissions.}
	\label{fig:transmitted_signal}
\end{figure}

\subsubsection{ZP-OFDM}\label{subsec:transmitted_signal_zp}
In this case, the $k$th transmitted symbol is defined as 
\begin{equation}
	\mathbf{x}_{{\rm zp}_k}[n] = \left\{\begin{matrix}
		\mathbf{x}_{{\rm zp}_k}[n] \sim \mathcal{CN}(0,\eta), &	0 \leq n \leq N_{\rm f}-1\\ 
		0, & N_{\rm f}\leq n < N-1
	\end{matrix}\right.,
\end{equation}
where $N_{\rm f}$ is the \ac{FFT} size and represents the number of data symbols per OFDM transmission, so that the first $N_{\rm f}$ samples are uncorrelated Gaussian, i.e., $\mathbb{E}(\mathbf{x}_{{\rm zp}_k}[n]\mathbf{x}_{{\rm zp}_k}^\dagger[m])=0$ for $n\neq m$. The last $N_{\rm zp}$ are zeroes that are padded to the signal. $N = N_{\rm f} + N_{\rm zp}$ is the number of samples per OFDM symbol and $\eta = {N/N_{\rm f}}$ such that $\mathbf{x}_{{\rm zp}_k}$ unitary power. 
The ZP transmission is depicted in the top graph of Fig.~\ref{fig:transmitted_signal}, where a continuous transmission is considered as
\begin{equation}\label{eq:x_zp}
\mathbf{x}_{\rm zp} = [\cdots \,\,\, \mathbf{x}_{{\rm zp}_{-2}}^{\rm T} \,\,\, \mathbf{x}_{{\rm zp}_{-1}}^{\rm T} \,\,\, \mathbf{x}_{{\rm zp}_0}^{\rm T} \,\,\, \mathbf{x}_{{\rm zp}_1}^{\rm T} \,\,\, \mathbf{x}_{{\rm zp}_2}^{\rm T} \cdots]^{\rm T}.
\end{equation}
\begin{figure}[t!]
	\centering
	\tikzstyle{block} = [draw, fill=white, rectangle, 
minimum height=1em, text width=7em,text centered]
\tikzstyle{block2} = [draw, fill=white, rectangle, 
minimum height=1em, minimum width=3em]

\tikzstyle{multiplier} = [draw,circle,fill=my_color!20,add={}{}{}{}] {} 
\tikzstyle{sum} = [draw,circle,scale=0.7,add2={}{}{}{}] {} 
\tikzstyle{input} = [coordinate]
\tikzstyle{output} = [coordinate]
\tikzstyle{pinstyle} = [pin edge={to-,thin,black}]
\newcommand\z{3}

\def\reference{(3,1)}

\def\windup{
	\tikz[remember picture,overlay]{
		\draw (-0.8,0) -- (0.8,0);
		\draw (-0.8,-0.4)--(-0.4,-0.4) -- (0.4,0.4) --(0.8,0.4);
}}


\centering
\begin{tikzpicture}[auto, node distance=0cm,>=latex',scale=1.0]
	\linespread{0.8} 
	
	
	\draw [] (-1.5,-3.25) rectangle (1.5,1.1);
	\node [] at (0,0.8) (tx) {\small DFRC BS};
	
	\node [block] at (0,0) (tx) {\small transmitter};
	\node [block] at (0,-1) (ic) {\small SI cancellation};
	\node [block,draw = my_color] at (0,-1.8) (cc) {\textcolor{my_color}{\small clutter cancellation}};
	\node [block] at (0,-2.8) (ed) {\small energy detection};
	
	\draw[->] (tx) -- (ic) node[yshift=0.50cm,xshift=0.35cm] {\small $\mathbf{x}[n]$};
	\draw[->] (ic) -- (cc) node[yshift=0.75cm,xshift=0.35cm] {};
	\draw[->] (cc) -- (ed) node[yshift=0.5cm,xshift=0.35cm] {\small $\mathbf{y}[n]$};
	
	\draw (tx) -| (1.75,.3) 	node[draw,isosceles triangle,anchor=west,rotate=90,   minimum height=0.2cm,minimum width =0.2cm,inner sep=0pt] (tx_antenna){};
	\draw (ic) -| (1.75,.3-1) 	node[draw,isosceles triangle,anchor=west,rotate=90,minimum size =0.2cm,inner sep=0pt](rx_antenna){};
	
	\node [] at (1.75,.7) (tx) {\small TX};
	\node [] at (1.75,-1.25) (tx) {\small RX};
	
	\node [block2] at (6,-0.35) (target) {\small target};
	\draw[-,color=applegreen,opacity = 0.7,dashed,line width=0.15mm] (1.85,.25) -- (5.45,-0.35) {};
	\draw[->,color=applegreen,opacity = 0.7,dashed,line width=0.15mm] (5.45,-0.35) -- (rx_antenna) {};
	\draw[->,color=azure(colorwheel),opacity = 0.7,dashed,line width=0.15mm] (1.85,.25) -- (1.85,.3-0.95){};

		\def\deltax{0.25}
		\def\deltay{-0.75}
		\draw[draw = my_color] (3.72,-1.92) ellipse (0.8cm and 0.4cm);
		\node at (3.72,-1.92) {\textcolor{my_color}{clutter}};
	 \draw[-,color=azure(colorwheel),opacity = 0.7,dashed,line width=0.15mm] (1.85,.25) -- (2.7+\deltax,-1.1+\deltay){};
	 \draw[->,color=azure(colorwheel),opacity = 0.7,dashed,line width=0.15mm] (2.7+\deltax,-1.1+\deltay) -- (1.85,.3-0.95){};
	 \draw[-,color=azure(colorwheel),opacity = 0.7,dashed,line width=0.15mm] (1.85,.25) -- (3.3+\deltax,-0.78+\deltay){};
	 \draw[->,color=azure(colorwheel),opacity = 0.7,dashed,line width=0.15mm] (3.3+\deltax,-0.78+\deltay) -- (1.85,.3-0.95){};
	 
	

\end{tikzpicture}
	\vspace{-0.3cm} 	
	\vspace{-0.7cm}
	\caption{\textcolor{my_color}{System model with focus on monostatic radar functionality with SI and clutter.}}
	\label{fig:system_model}
\end{figure}
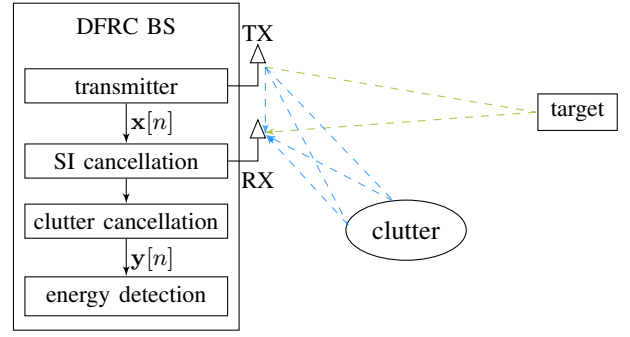

\subsubsection{CP-OFDM}\label{subsec:transmitted_signal_cp}
In this case, the transmitted signal is defined as $\mathbf{x}_{{\rm cp}_k}[n] \sim \mathcal{CN}(0,1)$ for $n=0,1,\cdots N-1$, where $N = N_{\rm f} + N_{\rm cp}$.
Since the CP is assumed to be inserted prior to transmission, we have $\mathbf{x}_{{\rm cp}_k}[n] = \mathbf{x}_{{\rm cp}_k}[m]$ for $m = n+N_{\rm f}$.
For completeness, we have
\begin{equation}
	\mathbf{x}_{{\rm cp}_k}[n] \sim \mathcal{N}(0,1), \,\,\,\,\,\,\, 0\leq n\leq N-1,
\end{equation}
where
\begin{equation}
	\mathbb{E}(\mathbf{x}_{{\rm cp}_k}[n]\mathbf{x}_{{\rm cp}_k}^\dagger[m]) = \left\{\begin{matrix}
		1, &	m=n \,\,\, \text{or} \,\,\, m = n + N_{\rm f} \\ 
		0, & \text{otherwise}
	\end{matrix}\right.,
\end{equation}
for $0\leq n,m \leq N-1$.
The CP transmission is represented in the bottom graph of Fig.~\ref{fig:transmitted_signal}, where a continuous transmission is considered as
\begin{equation}\label{eq:x_cp}
	\mathbf{x}_{\rm cp} = [\cdots \,\,\, \mathbf{x}_{{\rm cp}_{-2}}^{\rm T} \,\,\, \mathbf{x}_{{\rm cp}_{-1}}^{\rm T} \,\,\, \mathbf{x}_{{\rm cp}_0}^{\rm T} \,\,\, \mathbf{x}_{{\rm cp}_1}^{\rm T} \,\,\, \mathbf{x}_{{\rm cp}_2}^{\rm T} \cdots]^{\rm T}.
\end{equation}
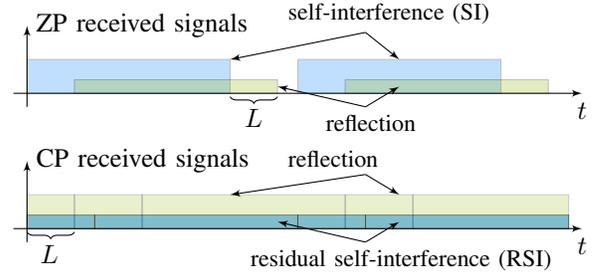
\begin{figure}[t]
	\centering
	\tikzstyle{block} = [draw, fill=white, rectangle, 
minimum height=3em, minimum width=3em]
\tikzstyle{block2} = [draw, fill=white, rectangle, 
minimum height=1em, minimum width=2.4em]

\tikzstyle{multiplier} = [draw,circle,fill=blue!20,add={}{}{}{}] {} 
\tikzstyle{sum} = [draw,circle,scale=0.7,add2={}{}{}{}] {} 
\tikzstyle{input} = [coordinate]
\tikzstyle{output} = [coordinate]
\tikzstyle{pinstyle} = [pin edge={to-,thin,black}]
\newcommand\z{3}

\def\windup{
	\tikz[remember picture,overlay]{
		\draw (-0.8,0) -- (0.8,0);
		\draw (-0.8,-0.4)--(-0.4,-0.4) -- (0.4,0.4) --(0.8,0.4);
}}


\centering
\begin{tikzpicture}[auto, node distance=0cm,>=latex',scale=0.9]
	
	\draw[->] (-0.2, 0) -- (8.2, 0) node[below] {$t$};
	\draw[->] (0, -0.2) -- (0, 1) node[right] {ZP received signals};
	\draw [fill=azure(colorwheel),nearly transparent] (0,0) rectangle (3,0.5);
	\draw [fill=azure(colorwheel),nearly transparent] (4,0) rectangle (7,0.5);
	
	\draw [fill=applegreen,nearly transparent] (0+0.7,0) rectangle (3+0.7,0.2);
	\draw [fill=applegreen,nearly transparent] (4+0.7,0) rectangle (7+0.7,0.2);
	
	\draw[->](5.0,0.9) -- (3.0,0.5){};
	\draw[->](5.0,0.9) -- (5.5,0.5) node[xshift=-1.6cm,yshift = 0.35cm,above right] {\small self-interference (SI)};
	
	\draw[->](5.0,-0.2) -- (3.7,.1){};
	\draw[->](5.0,-0.2) -- (5.5,0.1) node[xshift=-1.1cm,yshift = -0.25cm,below right] {\small reflection};

	
	\draw [decorate, decoration = {brace}] (3.7,-0.05) --  (3,-0.05);
%
	\node[below] at (3.35,-0.1) { $L$};

	\draw[->] (-0.2, 0-2) -- (8.2, 0-2) node[below] {$t$};
	\draw[->] (0, -0.2-2) -- (0, 1-2) node[right] {CP received signals};
	\draw [fill=azure(colorwheel),opacity = 0.5] (0,-2) rectangle (1,0.2-2);
	\draw [fill=azure(colorwheel),opacity = 0.5] (1,-2) rectangle (4,0.2-2);
	\draw [fill=azure(colorwheel),opacity = 0.5] (4,-2) rectangle (5,0.2-2);
	\draw [fill=azure(colorwheel),opacity = 0.5] (5,-2) rectangle (8,0.2-2);
	
	\draw [fill=applegreen,opacity = 0.2] (0+0.7,-2) rectangle (1+0.7,0.5-2);
	\draw [fill=applegreen,opacity = 0.2] (1+0.7,-2) rectangle (4+0.7,0.5-2);
	\draw [fill=applegreen,opacity = 0.2] (4+0.7,-2) rectangle (5+0.7,0.5-2);
	\draw [fill=applegreen,opacity = 0.2] (5+0.7,-2) rectangle (8+0.0,0.5-2);
	\draw [fill=applegreen,opacity = 0.2] (0,-2) rectangle (0.7,0.5-2);
	
		\draw[->](5.0,0.8-2) -- (3.0,0.5-2){};
	\draw[->](5.0,0.8-2) -- (5.5,0.5-2) node[xshift=-1.6cm,yshift = 0.25cm,above right] {\small reflection};
	
		\draw[->](5.0,-0.2-2) -- (3.7,.1-2){};
	\draw[->](5.0,-0.2-2) -- (5.5,0.1-2) node[xshift=-2.1cm,yshift = -0.25cm,below right] {\small residual self-interference (RSI)};

%
	\draw [decorate, decoration = {brace}] (0.7,-0.05-2) --  (0,-0.05-2);
	\node[below] at (0.33,-0.1-2) { $L$};
%
	
\end{tikzpicture}
	\vspace{-0.3cm}
	\vspace{-0.7cm}
	\caption{Received signal \textcolor{my_color}{without clutter} for transmission with ZP and CP. For the ZP system, only the silent periods are used for energy detection.}
	\label{fig:rx_signals}
\end{figure}

\subsection{Received Signal}
\textcolor{my_color}{In this work, we focus on the monostatic radar functionality of the \ac{DFRC} base station, which has SI and clutter cancellation blocks depicted in Fig.~\ref{fig:system_model}.
Thus, the radar channel is impaired by the RSI and \ac{RCI}.}

For the transmitted signal $\mathbf{x}[n]$ being either ZP or CP-OFDM defined \eqref{eq:x_zp} and \eqref{eq:x_cp}, respectively, the radar received signal after the interference cancellation block is given by
\begin{equation}\label{eq:yn1}
\begin{split}
	& H_0:  \mathbf{y}[n] = h_{\rm i} \mathbf{x}[n] + 	\textcolor{my_color}{\sum_{q=0}^{N_{\rm c}-1} h_{{\rm c}_q} \mathbf{x}[n-L_{{\rm c}_q}]} +  \mathbf{w}[n]
	\\ & H_1: \mathbf{y}[n] = h_{\rm i} \mathbf{x}[n] +	\textcolor{my_color}{\sum_{q=0}^{N_{\rm c}-1} h_{{\rm c}_q} \mathbf{x}[n-L_{{\rm c}_q}]} + \mathbf{w}[n]
	\\ & \,\,\,\,\,\,\,\,\,\,\,\,\,\,\,\,\,\,\,\,\,\,\,\,\,\,\,\,\,\,\,\, + h_{\rm t} \mathbf{x}[n-L_{\rm t}].
\end{split}
\end{equation}
The null hypothesis $H_0$ indicates that the target is absent meaning that there is no reflected signal from the target. 
\textcolor{my_color}{The coefficient $h_{\rm i} = \sqrt{\rho_{\rm si} \sigma^2} e^{2 \pi \phi_{\rm i}}$ represents the amplitude gain of the RSI signal due to the imperfect interference cancellation block, where $\rho_{\rm si}$ is the RSI power in relation to the \ac{AWGN} power $\sigma^2$ and $\phi_{\rm i}$ is the phase uniformly distributed between 0 and $2\pi$.
The coefficients $h_{{\rm c}_q} = \sqrt{\rho_{\rm ci}/N_{\rm c} \sigma^2} e^{2 \pi \phi_{{\rm c}_q}}$ represent the residual clutter gain for the $q$th clutter, where $\rho_{\rm ci}$ is also defined as the total clutter power in relation to $\sigma^2$, and $N_{\rm c}$ is the number of clutters such that $0 \leq q < N_{\rm c}$. 
It is implicitly assumed that all clutters have the same residual interference.
Lastly, the AWGN is defined as $\mathbf{w}[n] \sim \mathcal{CN}(0,\sigma^2)$.
As special cases,  $h_{\rm i}=0$ indicates perfect SI isolation, and  $h_{{\rm c}_q}=0 \forall q$ indicates either perfect clutter estimation or no clutter.}

\textcolor{my_color}{The hypothesis $H_1$ in \eqref{eq:yn1} indicated the presence of the target. In this case, $h_{\rm t} = |h_{\rm t}| e^{\phi_{\rm t}}$ is the gain of the reflected signal whose power is given by the well-known double path loss radar formula}
\begin{equation}\label{eq:hr}
	\textcolor{my_color}{|h_{\rm t}|^2 = \frac{P G^2(c/f)^2 \sigma_{\rm rcs}}{(4 \pi)^3 d^{2\alpha}} \,\,\,}
\end{equation}	
\textcolor{my_color}{where $P$ is the transmit power}, $G$ is the DFRC antenna gain which is assumed to be the same for the transmitter and receiver antennas, $f$ is the transmission frequency, $c$ is the speed of light, $\sigma_{\rm rcs}$ is the target's radar cross-section, $d$ is the distance between the DFRC and the target, and $\alpha$ is the path loss exponent.
Lastly, the discrete-time delay in \eqref{eq:yn1} is given by
\begin{equation}\label{eq:L}
	L_{\rm t} = {\rm round}\left(2d B/c\right),
\end{equation}
where ${\rm round}(\cdot)$ approximates its input to the nearest integer and $B$ is the signal sampling rate.
Under the $H_1$ hypothesis, the received signal for ZP and CP transmissions are depicted in Fig.~\ref{fig:rx_signals} for some  $L_{\rm t}$ smaller than $N_{\rm zp}$ and $N_{\rm cp}$.
In particular, we note that for the ZP transmission, $h_{\rm i}$ is irrelevant because only the samples for $\mathbf{x}[n]=0$ are taken by the energy detection module.
However, for the CP-OFDM it is desirable that $|h_{\rm i}|<|h_{\rm t}|$, indicating that the RSI power is small. 
It is also worth noticing that there is a minimum distance from which the ZP-OFDM is able to detect the reflection. 
Specifically, if $L_{\rm t}=0$ in \eqref{eq:L}, we see that there would be a reflected signal in Fig.~\ref{fig:rx_signals} during the ZP period since there would be no energy to be collected.
%


\subsection{Decision Variable and Performance Metrics}	
The decision variable is defined by the energy detection criterion as
\begin{equation}\label{eq:z}
	Z = \frac{1}{|I|}\sum_{n\in I}|\mathbf{y}[n]|^2,
\end{equation}	
where $I$ is the set containing the indexes of the samples whose power is averaged and $|I|$ is its cardinality.
The hypothesis testing towards the hypotheses $H_0$ and $H_1$ is performed by comparing $Z$ to a threshold $\lambda$ as
\begin{equation}\label{eq:hyp_testing}
Z \mathop{\lessgtr}_{H_1}^{H_0} \lambda,
\end{equation}
where the hypothesis $H_0$ is declared when $Z < \lambda$, otherwise $H_1$ is chosen.
According to \eqref{eq:hyp_testing}, the \ac{PFA} and \ac{PD} are respectively defined as
\begin{equation}
	P_{\rm fa} = {\rm Pr}(Z > \lambda| H_0)
%
\,\,\,\,\,\, \text{and}	\,\,\,\,\,\, P_{\rm d} = {\rm Pr}(Z > \lambda| H_1).
\end{equation} 
Typically, the threshold $\lambda$ is chosen to meet a certain \ac{CFAR}.

\textcolor{my_color}{Another performance metric is the \ac{KLD} between the PDFs $p_{Z| H_0}$ and $p_{Z| H_1}$, which is one type of statistical distance measurement.
The higher the KLD, the better the radar performance since this indicates that the hypotheses $H_0$ and $H_1$ are more discernible given the observation $Z$.
When $p_{Z| H_0} \sim \mathcal{N}(\mu_0,\sigma^2_0 )$ and $p_{Z| H_1} \sim \mathcal{N}(\mu_1,\sigma^2_1 )$ are Gaussian PDFs, the KLD is given by \cite{Liuxia}}
\begin{equation}\label{eq:kl_divergence_gaussian}
	\textcolor{my_color}{\begin{split}
		D(p_{Z| H_0}||p_{Z| H_1}) \!= \!\ln \left( \frac{\sigma_1}{\sigma_0}\right) \!+ \!\frac{\sigma_0^2 \!+ \!(\mu_0-\mu_1)^2}{2\sigma_1^2} \!- \!\frac{1}{2}.
	\end{split}}
\end{equation}

\section{Probability of Detection, ZP-OFDM}\label{sec:statistics_zp}
\textcolor{my_color}{In general, the energy of the reflected signal is computed for the samples within the range $N_{\rm f} + \Delta_{\rm s} < n \leq N-1$ as
	\begin{equation}\label{eq:z_zp}
		Z_{\rm zp} = \frac{1}{N_{\rm zp}-\Delta_{\rm s}}\sum_{n = N_{\rm f} + \Delta_{\rm s}}^{N-1}|\mathbf{y}[n]|^2,
	\end{equation}	
	where $-N_{\rm f} \leq \Delta_{\rm s} < N_{\rm zp}$}.
Notice that according to the definition of $\mathbf{x}_{\rm zp}$ in \eqref{eq:x_zp}, the samples $0 < n \leq N-1$ correspond to the OFDM symbol with index $k=0$, which is selected without loss of generality.
\subsection{Rejection of RSI and RCI}\label{subsec:clutter_rejection}
\textcolor{my_color}{We recall that in \cite{Sayed2}, the idea of using ZP-OFDM is to reject the RSI by taking the samples only during the silent periods, see Fig.~\ref{fig:rx_signals}, which corresponds to $\Delta_{\rm s}=0$ in \eqref{eq:z_zp}. 
	Nevertheless, we note that when there is a residual clutter with delay $0 < L_{{\rm c}_m} < N_{\rm zp}$, the radar signal is impaired similarly to RSI, but with a delayed interference.
	Thus, the ZP-OFDM system can reject the clutters that are closer than the target of interest by setting $\Delta_{\rm s}= L_{{\rm c}_{N_{\rm c}-1}}$ where $L_{{\rm c}_{N_{\rm c}-1}}$ is the highest delay of the clutters.
	In the case where $L_{{\rm c}_m} \geq N_{\rm zp}$, clutter rejection is not possible, and both CP and ZP-OFDM are equally impaired.}

\textcolor{my_color}{In the following we give special attention to the case $\Delta_{\rm s}=0$ without clutter because it avoids \ac{FD} operation at the radar unit by taking advantage of the silent periods.
	Subsequently, the general case of $\Delta_{\rm s}\neq 0$ is considered to investigate ZP-OFDM system under RSI and RCI.}

\subsection{\textcolor{my_color}{Special case: $\Delta_{\rm s}=0$ without clutter}}\label{subsec:special_case_delta0}
\subsubsection{sum-Gamma (exact)}
Under the hypothesis $H_1$, we can conveniently separate the decision variable $Z_{\rm zp}$ in \eqref{eq:z_zp} into two parts as
\begin{equation}\label{eq:Er}
	\begin{split}
		E_{\rm t}  = \frac{1}{N_{\rm zp}}\sum_{n = N_{\rm f}}^{N_{\rm f}+\-1}|h_{\rm t}\mathbf{x}_{\rm zp}[n-L_{\rm t}] + \mathbf{w}[n]|^2
	\end{split}
\end{equation}
%
%
\begin{equation}\label{eq:Ew}
	\hspace{-2.9cm}\text{ and} \qquad\qquad	E_{\rm w} = \frac{1}{N_{\rm zp}}\sum_{n = N_{\rm f}+\tilde{L}_{\rm t}}^{N-1}|\mathbf{w}[n]|^2,
\end{equation}
where $\tilde{L}_{\rm t} = \min(L_{\rm t},N_{\rm zp})$ to ensure a maximum window of $N_{\rm zp}$ samples. 
For the delay being limited to $0 < L_{\rm t} \leq N_{\rm f}$, one can verify that $Z_{\rm zp} = E_{\rm t} + E_{\rm w}$, where $E_{\rm t}$ contains the reflected signal plus noise, and $E_{\rm w}$ contains only noise.
\textcolor{my_color}{Based on the definition given in the Appendix \ref{apsubsec:gamma_statistics}, it is straightforward to verify that $E_{\rm t}\sim \Gamma(k_{\rm t},\theta_{\rm t})$ and $E_{\rm w}\sim \Gamma(k_{\rm w},\theta_{\rm w})$ follow the Gamma distribution with parameters} 
\begin{equation}\label{eq:E_r_param}
	k_{\rm t} = \tilde{L}_{\rm t} \,\,\,\,\,\, \text{and} \,\,\,\,\,\,\theta_{\rm t} = ( |h_{\rm t}|^2\eta+\sigma^2)/N_{\rm zp},
\end{equation}
and 
\begin{equation}\label{eq:E_w_param}
	k_{\rm w} = N_{\rm zp}-\tilde{L}_{\rm t} \,\,\,\,\,\, \text{and} \,\,\,\,\,\,\theta_{\rm w} = \sigma^2/N_{\rm zp},
\end{equation}
respectively.
There are two remarks to be made. 
Firstly, we note that if $N_{\rm f} < L_{\rm t} < N$, the delayed signal $\mathbf{x}_{\rm zp}[n-L_{\rm t}]$ contains zeroes\footnote{\textcolor{my_color}{This fact can be verified easily by assuming $N_{\rm f} < L_{\rm t} < N$ in Fig.~\ref{fig:rx_signals}.}}, which means that \eqref{eq:E_r_param} is valid for $0 < L_{\rm t} \leq N_{\rm f}$.
Secondly, if the target is so close such that $L_{\rm t}=0$, $E_{\rm t} = 0$ and therefore the PD is very small because $Z_{\rm zp}$ is the same for both hypotheses $H_0$ and $H_1$.
In practice, $L_{\rm t}=0$ can be avoided up to a certain limit by increasing the bandwidth of the transmitted signal, which increases the resolution and allows the detection of near targets.

Then, due to $Z_{\rm zp} = E_{\rm t} + E_{\rm w}$, $Z_{\rm zp}$ can be seen as the sum of two Gamma RVs in general, whose CDF is numerically computed as $F_{Z_{\rm zp}|H_1}(\lambda|k_{\rm t},\theta_{\rm t},k_{\rm w},\theta_{\rm w})$ in \eqref{eq:sum_gamma_cdf} in the Appendix~\ref{apsubsec:sum_gamma}.
A useful particular case of \eqref{eq:pd_zp_exact} is for $\tilde{L}_{\rm t}= N_{\rm zp}$, which means that $E_{\rm w} = 0$ because $k_{\rm w} = N_{\rm zp} - \tilde{L}_{\rm t} = 0$.
In this particular case, \eqref{eq:pd_zp_exact} can be written in terms of the Gamma CDF $F_{Z_{\rm zp}|H_1}(\lambda|k_{\rm t},\theta_{\rm t})$ in \eqref{eq:gamma_cdf}.
Then, the exact PD for the ZP system is given by
\begin{equation}\label{eq:pd_zp_exact}
	P_{{\rm d}_{\rm zp}} = \left\{\begin{matrix}
		1 - F_{Z_{\rm zp}|H_1}(\lambda_{\rm zp}|k_{\rm t},\theta_{\rm t},k_{\rm w},\theta_{\rm w}), &	\tilde{L}_{\rm t} < N_{\rm zp}\\ 
		1 - F_{Z_{\rm zp}|H_1}(\lambda_{\rm zp}|k_{\rm t},\theta_{\rm t}), &	\tilde{L}_{\rm t} = N_{\rm zp}
	\end{matrix}\right..
\end{equation}

\subsubsection{Approximation to Gamma}
For some set of parameters $(k_{\rm t},\theta_{\rm t})$ and $(k_{\rm w},\theta_{\rm w})$ when $\tilde{L}_{\rm t} < N_{\rm zp}$, computing \eqref{eq:pd_zp_exact} is numerically unstable.
As an alternative, we show that in some cases the PD of the ZP system is well approximated to a Gamma RV with parameters $(k_{\rm t},\theta_{\rm t})$ when $\theta_{\rm t}^2 k_{\rm t} \gg \theta_{\rm w}^2 k_{\rm w}$, which implies that $Z_{\rm zp} \approx E_{\rm t}$.
Then, the PD is written as
\begin{equation}\label{eq:pd_zp_gamma_approx}
	P_{\rm d_{zp}} \approx 1 - F_{E_{\rm t}|H_1}(\lambda_{\rm zp}|k_{\rm t},\theta_{\rm t})
\end{equation}
where $k_{\rm t}$ and $\theta_{\rm t}$ are given in \eqref{eq:E_r_param}.
The condition $\theta_{\rm t}^2 k_{\rm t} \gg \theta_{\rm w}^2 k_{\rm w}$ can be measured by
\begin{equation}\label{eq:sigma_R}
	\sigma_{\rm R} = \sqrt{\theta_{\rm t}^2 k_{\rm t}/(\theta_{\rm w}^2 k_{\rm w})},
\end{equation}
which is the ratio between the standard deviation of $E_{\rm t}$ and $E_{\rm w}$.
Basically, the higher the value of $\sigma_{\rm R}$, the more accurate \eqref{eq:pd_zp_gamma_approx} will be, which is observed in the numerical analysis of Section \ref{sec:numerical_results}.

\subsection{General case with Gaussian Approximation}\label{subsec:zp_gaussian_approx}
\textcolor{my_color}{In the general case of $\Delta_{\rm s}\neq 0$ and non-zero clutter, the distributions presented in the previous subsection do not describe $Z_{\rm zp}$ accurately.
	Under these circumstances, we can approximate the decision variable $Z_{\rm zp} \sim \mathcal{N}(\mu_{Z_{\rm zp}},\sigma_{Z_{\rm zp}} ^2)$ to Gaussian with moments $\mu_{Z_{\rm zp}} = \mathbb{E}(Z_{\rm zp})$ and $\sigma_{Z_{\rm zp}} ^2 = \mathbb{V}(Z_{\rm zp})$.
	By defining the variable
	\begin{equation}\label{eq:zn_zp}
		\begin{split}
			& z_{{\rm zp}_{n}} = |\mathbf{y}[n]|^2 = \\ & \left|h_{\rm i} \mathbf{x}_{\rm zp}[n] \!\!+ \!\!\sum_{q=0}^{N_{\rm c}\!-\!1}\! h_{{\rm c}_q} \mathbf{x}_{\rm zp}[n \!- \!L_{{\rm c}_q}] \!+\! \! h_{\rm t} \mathbf{x}_{\rm zp}[n-L_{\rm t}]\! \!+ \!\!\mathbf{w}[n]\right|^2,
		\end{split}
	\end{equation}
	the moments of $Z_{\rm zp}$ are written as
	\begin{equation}\label{eq:mu_zp}
		\mu_{Z_{\rm zp}} \!=  \! \frac{1}{N_{\rm zp}-\Delta_{\rm s}}\!\sum_{n = N_{\rm f} + \Delta_{\rm s}}^{N-1}\!\!\!\!\mathbb{E}(z_{{\rm zp}_{n}}) \!= \!\sum_{j = 0}^{N_{\rm c}+1}\!\dot{h}_j^2 C_{{\rm zp}_j} \!+ \!\sigma^2
	\end{equation}
	and
	\begin{equation}\label{eq:var_zp}
		\begin{split}
			\sigma_{Z_{\rm zp}} ^2 = & \frac{1}{(N_{\rm zp}-\Delta_{\rm s})^2}\sum_{n = N_{\rm f} + \Delta_{\rm s}}^{N-1}\mathbb{V}(z_{{\rm zp}_{n}}) 
			\\ & \frac{2}{(N_{\rm zp}-\Delta_{\rm s})^2}\sum_{n = N_{\rm f} + \Delta_{\rm s} }^{N-1}\sum_{m=n+1}^{N-1}\mathbb{C}(z_{{\rm zp}_{n}},z_{{\rm zp}_{m}}).
		\end{split}
	\end{equation}
	In \eqref{eq:mu_zp},
	\begin{equation}\label{eq:h_dot}
		\dot{h}_j^2 = \left\{\begin{matrix}
			|h_{\rm i}|^2, & j=0 \\
			|h_{{\rm c}_{j-1}}|^2, & 1 \leq j \leq N_{\rm c} \\
			|h_{\rm t}|^2, & j = N_{\rm c}+1 
		\end{matrix}\right.
	\end{equation}
	is a convenient definition that unifies the channel coefficients related to the SI, clutter, and target in the same variable, where its corresponding delay quantity is $\dot{L}_j$.}
\textcolor{my_color}{The computation of $\mathbb{E}(z_{{\rm zp}_{n}})$, $\mathbb{V}(z_{{\rm zp}_{n}})$ and $\mathbb{C}(z_{{\rm zp}_{n}},z_{{\rm zp}_{m}})$ along with some simplifications are described in the Appendix \ref{apsubsec:moments_ZP}.
	In particular, the rightmost side of \eqref{eq:mu_zp} is a simplification shown in \eqref{eq:mu_zp_simplification}, in which $C_{{\rm zp}_j} \in \mathbb{N}_0$ is given in \eqref{eq:C_zp_j}. 
	And equation \eqref{eq:cov_simplification} shows a simplification on how to compute the covariance terms in \eqref{eq:var_zp}.}

In summary, the PD approximation assuming that $Z_{\rm zp}~\sim~\mathcal{N}(\mu_{Z_{\rm zp}} ,\sigma_{Z_{\rm zp}}^2)$ is given by
\begin{equation}\label{eq:pd_zp_gauss_approx}
	P_{\rm d_{zp}} \approx 1 - \Phi_{Z_{\rm zp}|H_1}(\lambda_{\rm zp}|\mu_{Z_{\rm zp}} ,\sigma_{Z_{\rm zp}} ^2),
\end{equation}
where $\Phi_{Z_{\rm zp}}(\cdot|\mu_{Z_{\rm zp}},\sigma_{Z_{\rm zp}}^2)$ is the Gaussian CDF which is evaluated numerically.

\textcolor{my_color}{Lastly, we note that the approximated PFA is straightforward using equations \eqref{eq:mu_zp} and \eqref{eq:var_zp}.
	The only necessary modification is to set $\dot{h}_{N_{\rm c}+1}^2 = |h_{\rm t}|^2 = 0$ in \eqref{eq:h_dot}, which makes $z_{{\rm zp}_{n}} $ in \eqref{eq:zn_zp} fall under $H_0$.}

\section{Probability of Detection, CP-OFDM}\label{sec:statistics_cp}
\textcolor{my_color}{For the CP-OFDM transmission, the energy is computed over $N = N_{\rm f} + N_{\rm cp}$ samples with index $0 \leq n < N$ as
	\begin{equation}\label{eq:Z_cp}
		Z_{\rm cp} = \frac{1}{N}\sum_{n = 0}^{N-1}|\mathbf{y}[n]|^2.
	\end{equation}	
	This corresponds to the OFDM symbol with index $k=0$ according to \eqref{eq:x_cp} without loss of generality.
	For the CP-OFDM transmission, deriving the exact PDF of $Z_{\rm cp}$ is more complicated than the ZP-OFDM because of the SI terms that correlate the samples of $\mathbf{y}[n]$ over time. 
	Thus, in the following, we present Gamma and Gaussian-based approximations to compute the PD and PFA of CP-OFDM.}
\textcolor{my_color}{\subsection{Approximation to Gamma without Clutter}\label{subsec:cp_gamma_approx}
	Approximating $Z_{\rm cp}$ to Gamma has an important application of comparing the achieved range of ZP-OFDM and CP-OFDM, which will be explored in Section \ref{sec:range_comparison_CP_ZP}.
	In this case, we have
	\begin{equation}\label{eq:z_cp_n}
		z_{{\rm cp}_n}  = |\mathbf{y}[n]|^2 = |h_{\rm i}\mathbf{x}_{\rm cp}[n] + h_{\rm t}\mathbf{x}_{\rm cp}[n-L_{\rm t}] + \mathbf{w}[n]|^2,
	\end{equation}
	which is a Gamma RV with $k=1$ and $\theta = |h_{\rm i}|^2 + |h_{\rm t}|^2+\sigma^2$ according to \eqref{eq:gamma_moments}.
	Then, by neglecting the temporal correlation of \eqref{eq:z_cp_n}, i.e., assuming that $\mathbb{E}(z_n z_m) = \mathbb{E}(z_n) \mathbb{E}(z_m)$ for all $(n,m)$, the decision variable $Z_{\rm cp}$ in \eqref{eq:Z_cp} is given by
	\begin{equation}\label{eq:Pd_cp_approx}
		P_{\rm d_{cp}} \approx 1 - F_{Z_{\rm cp}|H_1}(	\lambda_{\rm cp}|k_{\rm cp},\theta_{\rm cp})
	\end{equation}
	with
	\begin{equation}\label{eq:Z_cp_gamma_approx}
		k_{\rm cp} = N  \,\,\,\,\,\, \text{and} \,\,\,\,\,\,\theta_{\rm cp} = (|h_{\rm i}|^2 + |h_{\rm t}|^2+\sigma^2)/N.
	\end{equation}
	Although a generalization to non-zero clutter is straightforward, the approximation becomes considerably less accurate as the new clutter terms introduce their correlation terms.
	\subsection{General case with Gaussian Approximation}\label{subsec:cp_gaussian_approx}
	For the CP-OFDM transmission, we have
	\begin{equation}\label{eq:zn_cp}
		\begin{split}
			& z_{{\rm cp}_n} = |\mathbf{y}[n]|^2 = \\ & \left|h_{\rm i} \mathbf{x}_{\rm cp}[n] \!\!+ \!\!\sum_{q=0}^{N_{\rm c}\!-\!1}\! h_{{\rm c}_q} \mathbf{x}_{\rm cp}[n \!- \!L_{{\rm c}_q}] \!+\! \! h_{\rm t} \mathbf{x}_{\rm cp}[n-L_{\rm t}]\! \!+ \!\!\mathbf{w}[n]\right|^2,
		\end{split}
	\end{equation}
	which is used to define the moments
	\begin{equation}\label{eq:mu_cp}
		\mu_{Z_{\rm cp}} =  \frac{1}{N}\sum_{n = 0}^{N-1}\mathbb{E}(z_{{\rm cp}_{n}}) = \sum_{j=0}^{N_{\rm c} + 1} \dot{h}_j^2  + \sigma^2,
	\end{equation}
	and
	\begin{equation}\label{eq:var_cp}
		\begin{split}
			\sigma_{Z_{\rm cp}} ^2 &= \frac{1}{N^2}\sum_{n = 0}^{N-1}\mathbb{V}(z_{{\rm zp}_{n}}) 
			+\frac{2}{N^2}\sum_{n = 0}^{N-1}\sum_{m=n+1}^{N-1}\mathbb{C}(z_{{\rm cp}_{n}},z_{{\rm cp}_{m}})
			\\ &= \frac{\mu_{Z_{\rm cp}}^2}{N}\!  + \!\frac{2}{N^2}\!\sum_{j=0}^{N_{\rm c}+1}\!\sum_{j'=j+1}^{N_{\rm c}+1}\!C_{{\rm cp}_{j,j'}}\dot{h}_j^2\dot{h}_{j'}^2\! + \! \frac{2}{N^2}\!\sum_{j=0}^{N_{\rm c}+1}\!C_{{\rm cp}_{j}}\dot{h}_j^4.
		\end{split}
	\end{equation}
	In \eqref{eq:mu_cp}, the rightmost part was found using the fact that the RSI, RCI and target have independent uniformly distributed phase, such that $\mathbb{E}(h_{\rm i} h_{{\rm c}_q}^\dagger) = \mathbb{E}(h_{\rm i} h_{{\rm t}}^\dagger) = \mathbb{E}(h_{{\rm c}_q} h_{{\rm t}}^\dagger) = \mathbb{E}(h_{{\rm c}_q} h_{{\rm c}_{q'}}^\dagger) = 0$ for all $q\neq q'$, where the quantity $\dot{h}_j^2$ is given by \eqref{eq:h_dot} to unify the RSI, clutter and target channels in the same variable.
	In the second line of \eqref{eq:var_cp}, $C_{{\rm cp}_{j,j'}}$ is given in \eqref{eq:C_cp_jj} and 
	\begin{figure*}[b]
		\textcolor{my_color}{	\hrulefill
			\begin{equation}\label{eq:C_cp_jj}
				C_{{\rm cp}_{j,j'}} = \min(N-\dot{L}_{j'},N_{\rm zp}) + \max(\min(N_{\rm f}-\dot{L}_{j'}),0) + 2\max(N_{\rm zp}-\dot{L}_{j'},0) + \max(\min(2(N_{\rm zp}-\dot{L}_j) + N_{\rm f} - \dot{L}_{j'},N_{\rm cp}-\dot{L}_j),0) + \dot{L}_j
		\end{equation}}
	\end{figure*}
	\begin{equation}\label{eq:C_cp_j}
		C_{{\rm cp}_{j}} = \max(N_{\rm cp}-\dot{L}_j,0) + \max(\dot{L}_j - N_{\rm f},0).
	\end{equation}
	The detailed demonstrations of \eqref{eq:C_cp_jj} and \eqref{eq:C_cp_j} are omitted due to the lack of space.
	Basically, $C_{{\rm cp}_{j,j'}}$ in \eqref{eq:C_cp_jj} accounts for the covariance terms between signals associated with $j$th and $j'$th channels of \eqref{eq:h_dot}, while $C_{{\rm cp}_{j}}$ considers the covariance term due to CP repetition within the signal associated with the $j$th channel.}
\textcolor{my_color}{Lastly, we note that we can manipulate the terms in \eqref{eq:var_cp} to quantify the contribution of the covariance terms in the variance as
	\begin{equation}\label{eq:C_tilde}
		\tilde{C} = 1-\frac{\mu_{Z_{\rm cp}}^2}{\sigma_{Z_{\rm cp}}^2 N},
	\end{equation}
	where $0\leq \tilde{C} \leq 1$. Basically, if $\tilde{C}$ increases, less accurate the Gamma approximation in \eqref{eq:Pd_cp_approx} is expected to be.}

For completeness, the PD approximation assuming that $Z_{\rm cp}~\sim~\mathcal{N}(\mu_{Z_{\rm cp}} ,\sigma_{Z_{\rm cp}}^2)$ is given by
\begin{equation}\label{eq:pd_cp_gauss_approx}
	P_{\rm d_{cp}} \approx 1 - \Phi_{Z_{\rm cp}|H_1}(\lambda_{\rm cp}|\mu_{Z_{\rm cp}} ,\sigma_{Z_{\rm cp}} ^2).
\end{equation}

\textcolor{my_color}{Computing the approximated PFA is straightforward using equations \eqref{eq:mu_cp} and \eqref{eq:var_cp}, where we should set $\dot{h}_{N_{\rm c}+1}^2 = |h_{\rm t}|^2 = 0$ such that $z_{{\rm cp}_n}$ in \eqref{eq:zn_cp} falls under $H_0$.}

\section{Upper Bound of ZP-OFDM PD without Clutter}\label{sec:pd_zp_upper_bound}
One practical use of the expressions derived in Section \ref{sec:statistics_zp} is to compute PD over distance to assess the achieved range detection for the ZP-OFDM transmission scheme.
As expected, different values of the ZP size $N_{\rm zp}$ lead to different PD over distance.
It turns out that for a given discrete-time delay $L_{\rm t}$ corresponding to the target's distance as \eqref{eq:L}, there is an optimal value of $N_{\rm zp}$ in terms of PD, which leads to a useful upper bound expression.
To demonstrate this finding, consider the following definition
\begin{definition}
	For a fixed system configuration of bandwidth, transmit power, frequency, antenna gain, and radar cross-section, we define $\tilde{P}_{\rm d_{zp}}(N_{\rm zp},L_{\rm t})$ as the probability of detection for a given ZP size $N_{\rm zp}$ and discrete-time delay $L_{\rm t}$.
\end{definition}
The above definition simply makes explicit the dependence of PD on $N_{\rm zp}$ and $L_{\rm t}$, while any other system parameter is fixed.
In the following, we claim that for the discrete-time delay $0<L_{\rm t}\leq N_{\rm f}$, the maximum of $\tilde{P}_{\rm d_{zp}}(N_{\rm zp},L_{\rm t})$ is achieved when $N_{\rm zp}=L_{\rm t}$.
Mathematically, the above claim is written as
\begin{equation}\label{eq:pd_max}
	\max_{N_{\rm zp}} \tilde{P}_{\rm d_{zp}}(N_{\rm zp},L_{\rm t}) = L_{\rm t},	
\end{equation}
for $0<L_{\rm t}\leq N_{\rm f}$, leading to the upper bound
\begin{equation}\label{eq:pd_ub}
	\tilde{P}_{\rm d_{zp}}^{\rm ub}(L_{\rm t}) = \tilde{P}_{\rm d_{zp}}(L_{\rm t},L_{\rm t}).
\end{equation}
In order to demonstrate that \eqref{eq:pd_max} holds, we first show that ${\tilde{P}_{\rm d_{zp}}(L_{\rm t},L_{\rm t})>\tilde{P}_{\rm d_{zp}}(L_{\rm t}+n,L_{\rm t})}$ holds for any positive integer $n$.
Then, we provide another argument for the same condition showing that ${\tilde{P}_{\rm d_{zp}}(L_{\rm t},L_{\rm t})>\tilde{P}_{\rm d_{zp}}(L-n,L)}$ holds for any positive integer $n$.
If these inequalities are valid, then \eqref{eq:pd_max} must be also valid. 
The interval $0<L_{\rm t}\leq N_{\rm f}$ is considered because it is the value of $L_{\rm t}$ for which the expressions of Section \ref{sec:statistics_zp} are valid.

\subsection{Showing that ${\tilde{P}_{\rm d_{zp}}(L_{\rm t},L_{\rm t})>\tilde{P}_{\rm d_{zp}}(L_{\rm t}+n,L_{\rm t})}$}
Recall that $Z_{\rm zp} = E_{\rm r} + E_{\rm w}$, see \eqref{eq:Er} and \eqref{eq:Ew}, where $E_{\rm r}$ contains the energy of the reflected signal plus noise, and $E_{\rm w}$ contains the solely the noise energy. $E_{\rm w} = 0$ for $N_{\rm zp} = L_{\rm t}$, meaning that the reflected signal spans the whole energy detection window of $N_{\rm zp}$ samples, which is the case considered by $\tilde{P}_{\rm d_{zp}}(L_{\rm t},L_{\rm t})$.
If this window is increased by $n$ samples as considered by $\tilde{P}_{\rm d_{zp}}(L_{\rm t}+n,L_{\rm t})$, there will be additional noise samples from $E_{\rm w}$, implying that $Z_{\rm zp}$ will be noisier concerning the previous case. 
Thus, ${\tilde{P}_{\rm d_{zp}}(L_{\rm t},L_{\rm t})>\tilde{P}_{\rm d_{zp}}(L_{\rm t}+n,L_{\rm t})}$ is a logical conclusion from the above observation that the decision variable is impaired in the second case.

\subsection{Showing that ${\tilde{P}_{\rm d_{zp}}(L_{\rm t},L_{\rm t})>\tilde{P}_{\rm d_{zp}}(L-n,L)}$}
For the case where $N_{\rm zp} = L-n$, for a positive integer $n$, we observe that $E_{\rm r}$ averages the energy of $L-n$ samples containing the reflected signal plus noise. 
Basically, if fewer samples than $L_{\rm t}$ are collected, the decision variable $Z_{\rm zp}$ will contain less information about the target than when all $L_{\rm t}$ samples are collected.
Again, ${\tilde{P}_{\rm d_{zp}}(L_{\rm t},L_{\rm t})>\tilde{P}_{\rm d_{zp}}(L-n,L)}$ is a logical conclusion from the above observation.

\subsection{Comments on the Upper Bound}
The results of \eqref{eq:pd_max} and \eqref{eq:pd_ub} have an important practical appeal.
One relevant aspect of the radar functionality of ISAC is to evaluate the range in which the DFRC can successfully detect targets.
Since the range can be expressed in terms of the discrete-time delay $L_{\rm t}$ according to \eqref{eq:L}, we can always have a maximum achieved detection probability $\tilde{P}_{\rm d_{zp}}^{\rm ub}(L)$ for a given distance by setting $N_{\rm zp}=L_{\rm t}$.
In practice, however, the zero-padding window size $N_{\rm zp}$ is fixed, meaning that the achievable probability of detection will deviate from its maximum.
Nevertheless, the upper bound  $\tilde{P}_{\rm d_{zp}}^{\rm ub}(L)$ is useful to measure the performance loss for a fixed $N_{\rm zp}$, which is a useful tool to help to choose $N_{\rm zp}$ in practical deployments.

\section{Range Comparison Between ZP and CP Systems}\label{sec:range_comparison_CP_ZP}
In general, we consider that the DFRC base station has an imperfect FD isolation, which is modeled by the RSI term $h_{\rm i}$ in \eqref{eq:yn1} and impacts negatively the detection performance of CP-OFDM.
Thus, it is relevant to have a metric that quantifies the performance loss due to imperfect FD isolation and compares the PD of both CP and ZP systems for different levels of RSI, which is the goal of this section.
\textcolor{my_color}{This analysis considers $\Delta_{\rm s} = 0$ in \eqref{eq:z_zp} and no clutter in order to evaluate solely the impact of RSI on the range.}

In order to accomplish the desired analysis, we first relate the distance to the SINRs of CP and ZP-OFDM systems.
Subsequently, the \ac{SINR} of CP and ZP-OFDM are related to their respective PD expressions, which allows a relation between their distances for a desired PD.
Thus, an expression comparing the range of both systems for a selected PD is given.

\subsection{Relating Distance to SINRs}

The radar \ac{SINR} of the CP system is given by 
\begin{equation}\label{eq:sinr_cp}
	\rho_{{\rm cp}} = \frac{|h_{\rm t}|^2}{|h_{\rm i}|^2  + \sigma^2}. 
\end{equation}
Since we are interested in providing a range analysis, in the following we combine the SINR \eqref{eq:sinr_cp} with the definition of $h_{\rm r}$ in \eqref{eq:hr}, such that the SINR and distance can be related as 
\begin{equation}\label{eq:d_cp}
	d_{\rm cp} = \left(\frac{1}{(\sigma^2+|h_{\rm i}|^2) \rho_{{\rm cp}}}\frac{PG^2(c/f)^2 \sigma_{\rm rcs}}{(4 \pi)^3}\right)^\frac{1}{2\alpha}. 
\end{equation}
For the ZP system, the SNR is given by
\begin{equation}\label{eq:rho_zp}
	\rho_{{\rm zp}} = \frac{|h_{\rm t}|^2 \eta}{\sigma^2},
\end{equation}
where $\eta = {N/N_{\rm f}}$ to ensure the same transmission power as the CP system. An analogous expression for the distance is given by
\begin{equation}\label{eq:d_zp}
	d_{\rm zp} = \left(\frac{\eta}{\sigma^2 \rho_{{\rm zp}}}\frac{P G^2(c/f)^2 \sigma_{\rm rcs}}{(4 \pi)^3}\right)^\frac{1}{2\alpha}. 
\end{equation}
\subsection{Relating PD to SNRs}
Now we related the SINRs to the PD formulas of Sections \ref{sec:statistics_zp} and \ref{sec:statistics_cp} according to the Appendix \ref{sec:ap_distance_zp} as
\begin{equation}\label{eq:rho_zp_PD}
	\rho_{{\rm zp}} \approx \frac{\lambda_{\rm norm_{\rm zp}}}{F^{-1}_{Z_{\rm zp}|H_1}(1-P_{\rm d}|N_{\rm zp},1)} - 1
\end{equation}	
\begin{equation}\label{eq:rho_cp_PD}
	\text{and} \,\,\,\,	\,\,\,\,\,\,\,\, \rho_{{\rm cp}} \approx \frac{N_{\rm f}\lambda_{\rm norm_{\rm cp}}}{F^{-1}_{Z_{\rm cp}|H_1}(1-P_{\rm d}|N_{\rm f},1)} - 1.
\end{equation}	
where $\lambda = F_G^{-1}(x|k,\theta)$ is the inverse Gamma CDF \eqref{eq:gamma_icdf} of Appendix \ref{apsubsec:gamma_statistics}.
The approximation of \eqref{eq:rho_zp_PD} is due to the acceptance of $\tilde{L} < N_{\rm zp}$ as \eqref{eq:pd_zp_gamma_approx}.
And the approximation \eqref{eq:rho_cp_PD} is due to the fact that the Gamma PD is an approximation for CP-OFDM, see \eqref{eq:Pd_cp_approx}.

\subsection{Distance Ratio of CP over ZP}
One crucial aspect of the inverse Gamma CDF is its property of allowing the scale parameter to be extracted as $F_G^{-1}(x|k,\theta) = F_G^{-1}(x|k,1)\theta$, see \eqref{eq:gamma_icdf} in the Appendix \ref{apsubsec:gamma_statistics}.
Because $\theta$ can be written in terms of the SINR, this allows writing the inverse PD in terms of the SINR, which in turn allows writing the distances \eqref{eq:d_cp} and \eqref{eq:d_zp} in the function of PD.
This can be done as long as the discrete-time delay for the ZP-OFDM transmission $L_{\rm zp} = {\rm round}\left({2d_{\rm zp} B}/{c}\right)$ is within the range $0<L_{\rm zp} \leq N_{\rm f}$ which is a condition in the PD expressions of Section \ref{sec:statistics_zp}, where $d_{\rm zp}$ in \eqref{eq:d_zp} is computed using $\rho_{\rm zp}$ in \eqref{eq:rho_zp_PD}.
Based on the previous considerations, the range comparison of CP and ZP-OFDM systems for a given PD is done via the metric
\begin{equation}\label{eq:delta}
	\begin{split}
		& \delta(\rho_{\rm si})   = \frac{d_{\rm cp}}{d_{\rm zp}} \approx \left(\frac{1}{\eta(1+\rho_{\rm si})} \frac{\rho_{\rm zp}}{\rho_{{\rm cp}}} \right)^{\frac{1}{2\alpha}}
		\\& \!\!\!= \textcolor{my_color}{\left(\frac{1}{\eta(1\!+\!\rho_{\rm si})}\frac{\lambda_{\rm norm_{\rm zp}}/F^{-1}_{Z_{\rm zp}|H_1}(1-P_{\rm d}|N_{\rm zp},1) \!- \!1}{\lambda_{\rm norm_{\rm cp}}/F^{-1}_{Z_{\rm cp}|H_1}(1\!-\! P_{\rm d}|N,1)\!-\!1}\right)^{\!\!\frac{1}{2\alpha}}}
	\end{split}
\end{equation}
for $0<L_{\rm zp} \leq N_{\rm f}$, where the SINRs $\rho_{{\rm zp}}$ and $\rho_{{\rm cp}}$ in equations \eqref{eq:rho_zp_PD} and \eqref{eq:rho_cp_PD} were replaced.
The term $\rho_{\rm si} = |h_{\rm i}|^2/\sigma^2$ is a convenient representation of the RSI level and represents the ratio between interference and noise power.

We highlight using \eqref{eq:rho_zp_PD} and \eqref{eq:rho_cp_PD} is necessary because we want to use the same PD for ZP and CP as a common factor for fairness.
Specifically, two useful analyses are possible.
We can analyze $\delta(0)$ to assess how larger the range of CP-OFDM is in relation to ZP-OFDM when a perfect FD cancellation is considered, implying that there is no RSI.
Another facet is to find the RSI $\rho_{\rm si}$ that makes the range of the CP and ZP systems to be the same, which is found at the point $\delta(\rho_{\rm si}) = 1$.
This can be easily written in closed form by rearranging \eqref{eq:delta}.
\textcolor{my_color}{Lastly, as shown in the Appendix \ref{sec:ap_distance_zp}, it is remarkable that \eqref{eq:delta} does not depend on system parameters such as power, frequency, bandwidth, etc, which makes the analysis general.}

\section{Numerical Results}\label{sec:numerical_results}

\begin{figure*}
	\input{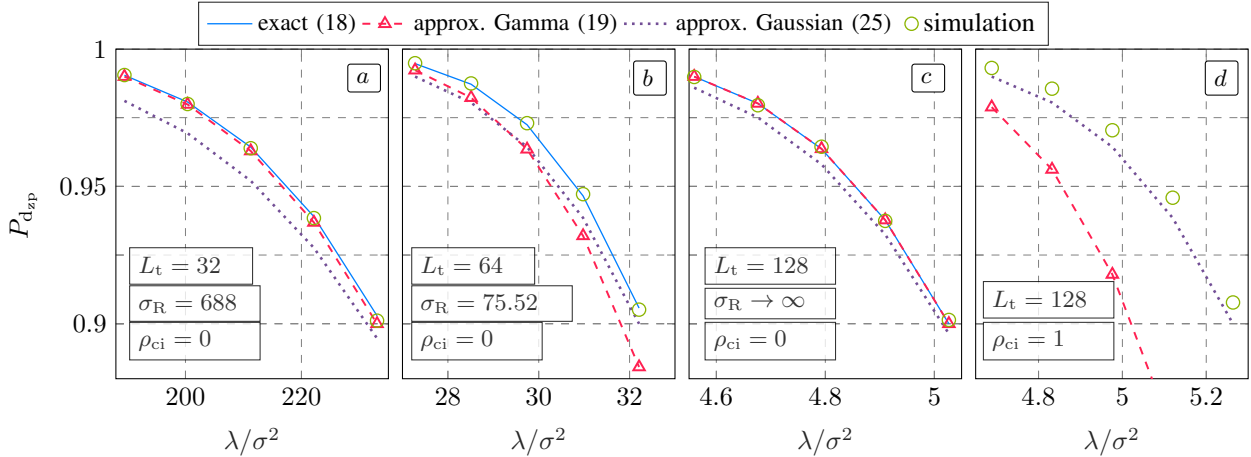} 
	\caption{\textcolor{my_color}{Validation of the theoretical PD curves derived in Section \ref{sec:statistics_zp} for the ZP system. The system parameters are given in Table \ref{tab:pd_validation}. Graphs $a$, $b$ and $c$ consider no clutter with $\rho_{\rm ci}=0$, and graph $d$ has clutter with $\rho_{\rm ci}=1$.}}
	\label{fig:pd_zp_validation}
\end{figure*}

\begin{figure*}
	\input{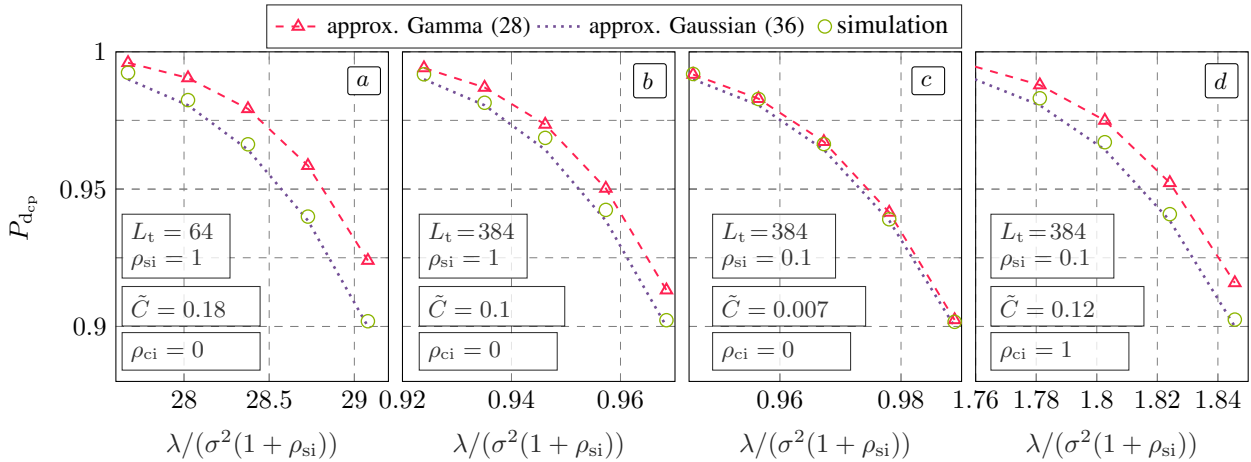} 
	\caption{\textcolor{my_color}{Validation of the theoretical PD curves derived in Section \ref{sec:statistics_cp} for the CP system. The system parameters are given in Table \ref{tab:pd_validation}. Graphs $a$, $b$ and $c$ consider no clutter with $\rho_{\rm ci}=0$, and graph $d$ has clutter with $\rho_{\rm ci}=1$.}}
	\label{fig:pd_cp_validation}
\end{figure*}

\subsection{Validation of Theoretical Expressions}\label{subsec:numerical_results_validation}
The goal of this subsection is to validate the theoretical PD formulas provided in Sections \ref{sec:statistics_zp} and \ref{sec:statistics_cp}, for the ZP and CP signals, respectively.
The system parameters are described in Table \ref{tab:pd_validation}. 
In all results, the path loss exponent in \eqref{eq:hr} is $\alpha = 2$ and the noise power is computed as $\sigma^2 = B\cdot N_{\rm psd}$, where ${N_{\rm psd} = 10^{-3}\cdot 10^{-174/10}}$ for a noise power spectrum density (PSD) of $-174 \,{\rm dBm}$.

\begin{table}[t!]
	\centering
	\caption{Simulation Parameters for Fig.~\ref{fig:pd_zp_validation} and Fig.~\ref{fig:pd_cp_validation}.}	\vspace{-0.3cm}
	\begin{tabular}{lc|lc}
		\toprule
		Parameter		& value & Parameter & value \\
		\midrule
		frequency, $f_c$	& $\SI{2.4} \GHz$ & ZP/CP size, $N_{\rm zp},N_{\rm cp}$ & 128  \\
		bandwidth, $B$	& $\SI{100} \MHz$ 	& \textcolor{my_color}{ZP sample shift, $\Delta_{\rm s}$} & \textcolor{my_color}{0}			\\
		antenna gain, $G$	& $16$ 		& 	\textcolor{my_color}{clutter echos, $N_{\rm c}$} & \textcolor{my_color}{$\left\{0,1\right\}$}		\\
		radar cross-sec., $\sigma_{\rm RCS}$		&$\SI{10}\m^2$ & 	\textcolor{my_color}{clutter delay, $L_{{\rm c}_0}$} & \textcolor{my_color}{$32$}  \\
		transmit power, $P$	& $20 {\rm dBm}$  & 	\textcolor{my_color}{RCI power, $\rho_{\rm ci}$ } & \textcolor{my_color}{$\left\{0,1\right\}$} \\
		FFT size, $N_{\rm f}$  & 512 \\ 
		\bottomrule
	\end{tabular}
	\label{tab:pd_validation}
	\vspace{-0.3cm}
\end{table}

\begin{figure*}
	\input{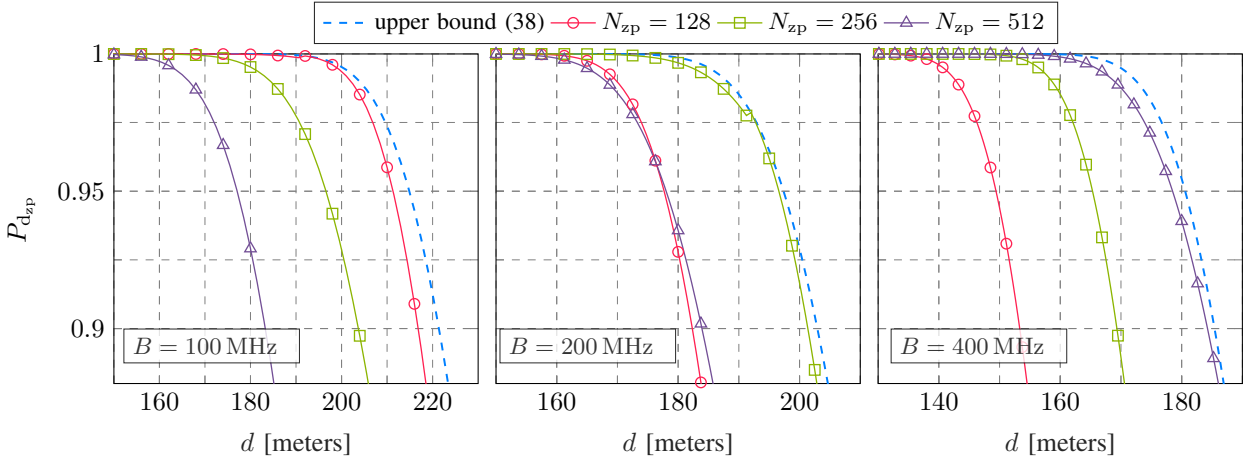} 
	\caption{PD of ZP system for different $N_{\rm zp}$ and bandwidth. The upper bound \eqref{eq:pd_ub} is plotted for comparison.}
	\label{fig:pd_ub2}
\end{figure*}

The results for the ZP system are depicted in Fig.~\ref{fig:pd_zp_validation}, \textcolor{my_color}{where graphs $a$, $b$ and $c$ consider no clutter with $\rho_{\rm ci}=0$, and graph $d$ has clutter with $\rho_{\rm ci}=1$.} 
The first observation we make is to note that the exact PD given in equation \eqref{eq:pd_zp_exact} matches the simulation accurately in graphs with no clutter $a$, $b$, and $c$, showing that it is correct.
In addition, we highlight that we have computed the variable $\sigma_{\rm R}$ of \eqref{eq:sigma_R} in order to show how this quantity is related to the accuracy of the approximate Gamma and Gaussian expressions of \eqref{eq:pd_zp_gamma_approx} and \eqref{eq:pd_zp_gauss_approx}, respectively.
\textcolor{my_color}{In the graph $a$} with $L_{\rm t}=32$ and $\sigma_{\rm R} = 688$, we see that the Gamma expressions lead to a much more accurate approximation than the Gaussian.
The reason is that the first $L_{\rm t}=32$ samples which are averaged have considerably more power than the remaining $N_{\rm zp}-{L_{\rm t}} = 96$ noise-only samples, which is measured by the relatively high value of $\sigma_{\rm R} = 688$, indicating that the noise only samples have little contribution to the decision variable $Z_{\rm zp}$ in \eqref{eq:z_zp}.
\textcolor{my_color}{In the graph $b$}, with $L_{\rm t}=64$ and $\sigma_{\rm R} = 75.52$, we see that neglecting the noise-only samples leads to a noticeable discrepancy when the Gamma approximation is used for $P_{\rm zp} \approx 0.9$, which is a typical value of interest. In this case, the Gaussian approximation is more appropriate. 
For smaller values of $\sigma_{\rm R}$, the gap tends to increase.
\textcolor{my_color}{In the graph $c$}, for $L_{\rm t}=N_{\rm zp}=128$, the value $\sigma_{\rm R}$ tends to infinity because there are no more noise-only samples given that the reflected signal occupies the whole zero-padding window.
In this case, the Gamma distribution becomes exact as stated in \eqref{eq:pd_zp_exact}. 
Since in some cases the exact PD \eqref{eq:pd_zp_exact} is not numerically stable, the value $\sigma_{\rm R}$ can be used to choose between the Gaussian or Gamma approximations.
In our numerical analysis, we have observed that typically for $\sigma_{\rm R}>300$ the Gamma approximation is more accurate.
\textcolor{my_color}{Lastly, graph $d$ shows that the Gamma approximation becomes poor under clutter conditions, which is expected since it does not take the clutter into account. 
	For this reason, the Gaussian approximation will be used to assess the system with clutter and the general setting of $\Delta_{\rm s} \neq 0$.}

The results related to the CP system are shown in Fig.~\ref{fig:pd_cp_validation}.
In particular, we are interested in assessing how the system parameters impact the accuracy of the approximate PD curves of \eqref{eq:Pd_cp_approx} and \eqref{eq:pd_cp_gauss_approx}.
As we have discussed in Section \ref{sec:statistics_cp}, the normalized covariance quantity $\tilde{C}$ in \eqref{eq:C_tilde} can be used to determine how much of the variance of the decision variable $Z_{\rm cp}$ in \eqref{eq:Z_cp} is due to temporal correlation between the RSI, CSI and reflection. 
\textcolor{my_color}{In the graph $a$, for $L_{\rm t}=64$ and $\rho_{\rm si}=1$, we have $\tilde{C}=0.18$ which is relatively large.
	As the result shows, the Gamma approximation is considerably less accurate than the Gaussian.
	In the graph $b$, $L$ is increased to 384, which leads to $\tilde{C}=0.1$.
	Clearly, the decrease of $\tilde{C}$ improves the accuracy of the Gamma approximation, but still, the Gaussian approximation is better.}
This result is interesting because it shows that the Gamma approximation does not depend only on $\rho_{\rm si}$.
Obviously, for smaller values of $\rho_{\rm si}$, e.g., 0.1 in the \textcolor{my_color}{graph $c$}, $\tilde{C}$ is decreased to 0.007, which leads to a good approximation of the Gamma-based PD.
The results show that for $\tilde{C}<0.01$, the Gamma approximation fits well the data.
\textcolor{my_color}{Lastly, graph $d$ exploits the effect of clutter in the PD by setting $\rho_{\rm ci} = 1$. In this case, $\tilde{C}$ jumps to 0.12 which basically has a very similar effect as having higher RSI as $\rho_{\rm si} = 1$.
	Clearly, the Gaussian approximation fits best in this case.}

\subsection{Upper Bound and $N_{\rm zp}$ Analysis}\label{subsec:numerical_results_upper_bound}
This subsection has the objective of demonstrating the usefulness of the analysis provided in Section \ref{sec:pd_zp_upper_bound}. 
The system parameters for the subsequent analysis are described in \mbox{Table \ref{tab:pd_upper_bound}}.

\begin{table}[t!]
	\centering
	\caption{Simulation Parameters for Fig.~\ref{fig:pd_ub2}.}	\vspace{-0.3cm}
	\begin{tabular}{lccc}
		\toprule
		Param.		& value  \\
		\midrule
		frequency, $f_c$	& $\SI{24} \GHz$ \\
		bandwidth $B$	& $\left\{{100},{200},{400}\right\} \SI{}\MHz$ 				\\
		antenna gain, $G$	& $16$ 				\\
		radar cross-section, $\sigma_{\rm RCS}$		&$\SI{10}\m^2$  \\
		transmit power, $P$	& $20 {\rm dBm}$  \\
		FFT size, $N_{\rm f}$  & $4\cdot N_{\rm zp}$ \\ 
		ZP size, $N_{\rm zp}$ & $\left\{128,256,512\right\}$ \\
		\bottomrule
	\end{tabular}
	\label{tab:pd_upper_bound}
\end{table}

Since the discrete-time delay $L_{\rm t}$ is associated with distance according to \eqref{eq:L} and the ZP size $N_{\rm zp}$ is fixed, we can use the upper bound as the highest possible for a given delay $L_{\rm t}$.
This analysis is shown in Fig.~\ref{fig:pd_ub2}, where three values for bandwidth are considered $B \in \left\{{100},{200},{400}\right\}\SI{}\MHz$.
The results consider $N_{\rm zp}$ can assume 128, 256, and 512.
The first consideration to make is that, indeed, the upper bound always leads to higher PD for all curves for fixed $N_{\rm zp}$.
In the left-most graph, with $B=\SI{100}{\MHz}$, we see that the system with $N_{\rm zp}=128$ achieves the highest PD, which is very close to the upper bound.
Interestingly, this result reveals that a bad choice of $N_{\rm zp}$ can cause significant performance loss in terms of range.
In particular, setting $N_{\rm zp} = 512$ for the system with $B=\SI{100}{\MHz}$ decreases the detection range from 220 to 183 meters at $P_{\rm d}\approx 0.9$.
When the bandwidth is increased, e.g., to $\SI{200}{\MHz}$ in the middle graph, it is observed that the system with $N_{\rm zp} = 256$ is now close to the upper bound, while the other configurations have a smaller range.
This can be explained by the fact that with double the bandwidth, it is necessary to collect twice as many samples within the same amount of time.
This trend is observed further in the rightmost graph for $\SI{400}{\MHz}$, where $N_{\rm zp} = 512$ is closer to the upper bound.

In summary, this analysis investigated how the numerology of the communication system is related to the detection performance of ZP systems.

\subsection{Detection Range Comparison Between CP and ZP System}\label{subsec:numerical_results_range_comparison}
The goal of this subsection is to compare the ZP and CP systems in terms of detection range using the expression $\delta(\rho_{\rm si})  = {d_{\rm cp}}/{d_{\rm zp}}$ in \eqref{eq:delta}, developed in Section \ref{sec:range_comparison_CP_ZP}.
We consider the RSI in decibel scale as $\tilde{\rho}_{\rm si} = 10 \log_{10} \rho_{\rm si}$.
In order to show the generality of the results, we consider three configuration setups given in Table~\ref{tab:range_comparison}.

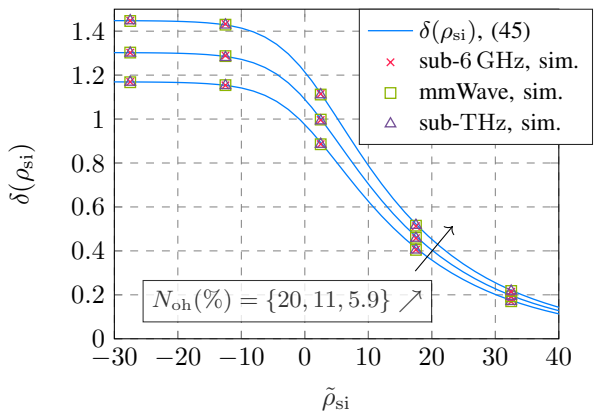
\begin{figure}[t!]
	\centering
%
%
\definecolor{mycolor1}{rgb}{0.00000,0.44700,0.74100}%
\definecolor{mycolor2}{rgb}{0.85000,0.32500,0.09800}%
\definecolor{mycolor3}{rgb}{0.92900,0.69400,0.12500}%
\definecolor{mycolor4}{rgb}{0.49400,0.18400,0.55600}%
\begin{tikzpicture}
	\begin{axis}[%
		scale = 0.6,
		width=4.5in,
		height=3.5in,
		at={(0in,0in)},
		xmin=-30,
		xmax=40,
		xlabel style={font=\color{white!15!black}},
		xlabel={$\tilde{\rho}_{\rm si}$},
		ylabel = {$\delta({\rho}_{\rm si})$},
		ymin=0.0,
		ymax=1.5,
		yminorticks=true,
		grid = both,
		ytick = {0,0.2,0.4,...,2},
		xtick = {-40,-30,-20,...,40},
		major grid style = {
			dashed,
			gray,
			line width = 0.25,
		},
		major tick style = {																			
			line width = 0.25pt,
			major tick length = 3pt,
		},		
		minor grid style = {
			dashed,
			gray,	
			line width = 0.25,
		},
		minor tick style = {																			
			line width = 0.25pt,
			minor tick length = 3pt,																																		
		},
		legend style={at={(0.55,0.588)}, anchor=south west, legend cell align=left, align=left, draw=white!15!black,legend columns=1}
		]
		
		
		\addplot [azure(colorwheel), line width=0.5pt]
table[row sep=crcr]{
	-37.5   1.1693502\\-35.663265   1.1693228\\-33.826531   1.1692811\\-31.989796   1.1692173\\-30.153061   1.1691201\\-28.316327   1.1689717\\-26.479592   1.1687453\\-24.642857   1.1684001\\-22.806122   1.1678739\\-20.969388   1.1670725\\-19.132653   1.1658534\\-17.295918   1.164002\\-15.459184   1.1611982\\-13.622449   1.1569692\\-11.785714   1.1506308\\-9.9489796   1.1412205\\-8.1122449   1.1274473\\-6.2755102   1.1077083\\-4.4387755   1.0802577\\-2.6020408   1.0436128\\-0.76530612   0.99715592\\1.0714286   0.94164758\\2.9081633   0.87924572\\4.744898   0.81292709\\6.5816327   0.74567978\\8.4183673   0.6799224\\10.255102   0.61730441\\12.091837   0.55877607\\13.928571   0.5047635\\15.765306   0.4553448\\17.602041   0.41038769\\19.438776   0.36964447\\21.27551   0.33281263\\23.112245   0.2995716\\24.94898   0.26960383\\26.785714   0.24260625\\28.622449   0.21829581\\30.459184   0.19641178\\32.295918   0.17671594\\34.132653   0.1589918\\35.969388   0.14304338\\37.806122   0.12869357\\39.642857   0.11578262\\41.479592   0.10416654\\43.316327   0.093715613\\45.153061   0.084313081\\46.989796   0.075853826\\48.826531   0.068243252\\50.663265   0.061396234\\52.5   0.055236178\\
};
		\addlegendentry{\small $\delta(\rho_{\rm si})$, \eqref{eq:delta}};
	
		\addplot [awesome, line width=.5pt,mark=x,only marks]
table[row sep=crcr]{
	-27.5   1.1692913\\-12.5   1.1535433\\2.5   0.88582677\\17.5   0.40551181\\32.5   0.16929134\\
};
		\addlegendentry{\small sub-$\SI{6}{\GHz}$, sim.};
	
			\addplot [applegreen, line width=.5pt,mark=square,only marks]
table[row sep=crcr]{
	-27.5   1.1675854\\-12.5   1.154355\\2.5   0.88533627\\17.5   0.40463065\\32.5   0.17089305\\
};

\addlegendentry{\small mmWave, sim.};
	
	\addplot [darklavender, line width=.5pt,mark=triangle,only marks,mark size = 2.3]
table[row sep=crcr]{
	-27.5   1.1678487\\-12.5   1.1524823\\2.5   0.88534279\\17.5   0.40425532\\32.5   0.17021277\\
};
\addlegendentry{\small sub-$\SI{}{\THz}$, sim.};
		
		\addplot [azure(colorwheel), line width=0.5pt]
table[row sep=crcr]{
	-37.5   1.3025674\\-35.663265   1.3025369\\-33.826531   1.3024904\\-31.989796   1.3024194\\-30.153061   1.3023111\\-28.316327   1.3021458\\-26.479592   1.3018937\\-24.642857   1.3015093\\-22.806122   1.3009235\\-20.969388   1.3000315\\-19.132653   1.2986752\\-17.295918   1.2966168\\-15.459184   1.2935022\\-13.622449   1.288811\\-11.785714   1.2817932\\-9.9489796   1.2714015\\-8.1122449   1.2562427\\-6.2755102   1.2346027\\-4.4387755   1.204627\\-2.6020408   1.1647319\\-0.76530612   1.1142055\\1.0714286   1.0537346\\2.9081633   0.98547442\\4.744898   0.91252691\\6.5816327   0.83812943\\8.4183673   0.76500469\\10.255102   0.69508214\\12.091837   0.62952295\\13.928571   0.56888717\\15.765306   0.51332269\\17.602041   0.46272138\\19.438776   0.41683052\\21.27551   0.37532565\\23.112245   0.33785547\\24.94898   0.30406804\\26.785714   0.27362524\\28.622449   0.24621006\\30.459184   0.22152972\\32.295918   0.19931631\\34.132653   0.17932615\\35.969388   0.16133842\\37.806122   0.14515354\\39.642857   0.13059142\\41.479592   0.11748971\\43.316327   0.10570214\\45.153061   0.09509702\\46.989796   0.085555815\\48.826531   0.07697183\\50.663265   0.069249058\\52.5   0.06230111\\
};
	
			\addplot [awesome, line width=.5pt,mark=x,only marks]
table[row sep=crcr]{
	-27.5   1.3022222\\-12.5   1.2888889\\2.5   0.99555556\\17.5   0.45777778\\32.5   0.19111111\\
};
		
		\addplot [applegreen, line width=.5pt,mark=square,only marks]
table[row sep=crcr]{
	-27.5   1.3026152\\-12.5   1.2864259\\2.5   0.99750934\\17.5   0.45952677\\32.5   0.19302615\\
};
		
\addplot [darklavender, line width=.5pt,mark=triangle,only marks,mark size = 2.3]
table[row sep=crcr]{
	-27.5   1.3004005\\-12.5   1.2803738\\2.5   0.99198932\\17.5   0.45794393\\32.5   0.19359146\\
};


\addplot [azure(colorwheel), line width=0.5pt]
table[row sep=crcr]{
	-37.5   1.4481892\\-35.663265   1.4481553\\-33.826531   1.4481036\\-31.989796   1.4480246\\-30.153061   1.4479042\\-28.316327   1.4477205\\-26.479592   1.4474402\\-24.642857   1.4470129\\-22.806122   1.4463619\\-20.969388   1.4453707\\-19.132653   1.4438638\\-17.295918   1.4415778\\-15.459184   1.4381207\\-13.622449   1.4329179\\-11.785714   1.4251435\\-9.9489796   1.4136493\\-8.1122449   1.3969163\\-6.2755102   1.3730863\\-4.4387755   1.3401588\\-2.6020408   1.2964243\\-0.76530612   1.2410858\\1.0714286   1.1748098\\2.9081633   1.0998262\\4.744898   1.0194262\\6.5816327   0.93712824\\8.4183673   0.85596485\\10.255102   0.77813931\\12.091837   0.70501481\\13.928571   0.63727726\\15.765306   0.57513766\\17.602041   0.51850649\\19.438776   0.46712134\\21.27551   0.42063174\\23.112245   0.37865207\\24.94898   0.3407928\\26.785714   0.30667794\\28.622449   0.27595395\\30.459184   0.24829371\\32.295918   0.22339758\\34.132653   0.2009928\\35.969388   0.18083208\\37.806122   0.16269188\\39.642857   0.14637039\\41.479592   0.1316857\\43.316327   0.11847391\\45.153061   0.10658742\\46.989796   0.095893387\\48.826531   0.086272222\\50.663265   0.077616324\\52.5   0.069828868\\
};

\addplot [awesome, line width=.5pt,mark=x,only marks]
table[row sep=crcr]{
	-27.5   1.4477612\\-12.5   1.4328358\\2.5   1.1144279\\17.5   0.51243781\\32.5   0.21393035\\
};
\addplot [applegreen, line width=.5pt,mark=square,only marks]
table[row sep=crcr]{
	-27.5   1.446304\\-12.5   1.4295676\\2.5   1.111576\\17.5   0.51464435\\32.5   0.21757322\\
};
%
\addplot [darklavender, line width=.5pt,mark=triangle,only marks,mark size = 2.3]
table[row sep=crcr]{
	-27.5   1.4476048\\-12.5   1.4281437\\2.5   1.1107784\\17.5   0.51646707\\32.5   0.21706587\\
};

	\end{axis}
	
	\draw[->] (4,0.9)--(4.5,1.5){};
		
		\draw[rounded corners = 0pt	] (2.3,0.5)
		node[fill = white, opacity = 0.8,draw=black, 
		]{\textcolor{my_color}{\small  $N_{\rm oh}(\%) = \left\{ 20, 11, 5.9\right\} \nearrow$}};
	%
	%

\end{tikzpicture}%
\vspace{-0.3cm}
	\caption{Range comparison analysis $d_{\rm cp}/d_{\rm zp}$ for $P_{\rm d} = 0.9$. Curves for different percentages \textcolor{my_color}{$N_{\rm oh} = N_{\rm zp}/(N_{\rm f}+N_{\rm zp})$} are shown. Simulations consider $10 \log_{10} {\rho}_{\rm si} = \left\{-27.5,-12.5,2.5,17.5\right\}$.}
	\label{fig:distance_ratio}
\end{figure}

The distance ratio expression \eqref{eq:delta} is evaluated in Fig.~\ref{fig:distance_ratio}, which considers different values of $N_{\rm zp}$ as $64$, $128$ and $256$. 
\textcolor{my_color}{We highlight that the following results are more meaningful to be described in terms of the ZP overhead $N_{\rm oh}(\%)~=~100N_{\rm zp}/(N_{\rm f}~+~N_{\rm zp})$ instead of $N_{\rm zp}$, since scaling both $N_{\rm zp}$ and $N_{\rm f}$ leads to a similar distance ratio.
For $N_{\rm f} = 1024$, and the above ZP sizes of $N_{\rm zp}$ as $64$, $128$ and $256$, the ZP overhead is $N_{\rm oh} = 5.9\%$, $11\%$ and $20\%$.}
Additionally, we plot the simulated values of $d_{\rm cp}/d_{\rm zp}$ for $P_{\rm d} = 0.9$ for some selected values of $\rho_{\rm si}$ to asses th accuracy of $\delta(\rho_{\rm si})$.
We observe that the simulated distance ratio values meet the theoretical expression for all tested $\rho_{\rm si}$, which indicates its usefulness.
Also, we highlight that the ZP-OFDM discrete-time delay range $0<L_{\rm zp} \leq N_{\rm f}$ is respected for $P_{\rm zp} = 0.9$.

One particular point of interest in the function $\delta(\rho_{\rm si})$ is its maximum value for $\rho_{\rm si} = 0$, or $\tilde{\rho}_{\rm si}\to -\infty$.
It turns out that $\delta(0)$ equals $1.45$, $1.3$ and $1.17$ for $N_{\rm oh}$ being $5.9\%$, $11\%$ and $20\%$, respectively.
As expected, the smaller $N_{\rm zp}$ in relation to the OFDM block size, the higher the performance loss of the ZP system in relation to CP under perfect FD isolation.
Interestingly, it is worth commenting that for the common value of $N_{\rm oh}=20\%$, the range of the CP system is only $17\%$ larger if perfect FD isolation is employed.
This indicates that depending on the costs and complexity of implementing the FD cancellation for the CP system, it may be more advantageous to reduce its detection range by using the ZP system to have a simpler setup.

\begin{figure}[t!]
	\centering
%
%
\definecolor{mycolor1}{rgb}{0.00000,0.44700,0.74100}%
\definecolor{mycolor2}{rgb}{0.85000,0.32500,0.09800}%
\definecolor{mycolor3}{rgb}{0.92900,0.69400,0.12500}%
\definecolor{mycolor4}{rgb}{0.49400,0.18400,0.55600}%
	\begin{tikzpicture}
	\begin{axis}[%
		scale = 0.6,
		width=3.7in,
		height=3.5in,
		at={(4.5in,0in)},
		xmin=105,
		xmax=160,
		xlabel style={font=\color{white!15!black}},
		xlabel={$d$, [meters]},
		ylabel = {$P_{\rm d}$},
		ymin=0.88,
		ymax=1,
		yminorticks=true,
		ylabel near ticks,
		grid = both,
		xtick = {100,110,120,...,160},
		minor xtick = {100,105,110,...,160},
				minor ytick = {0.8,.825,0.850,...,1},
		major grid style = {
			dashed,
			gray,
			line width = 0.25,
		},
		major tick style = {																			
			line width = 0.25pt,
			major tick length = 3pt,
		},		
		minor grid style = {
			dashed,
			gray,	
			line width = 0.25,
		},
		minor tick style = {																			
			line width = 0.25pt,
			minor tick length = 3pt,																																		
		},
		legend style={at={(0.95,0.408)}, anchor=south west, legend cell align=left, align=left, draw=white!15!black,legend columns=1}
		]

		\addplot [azure(colorwheel), line width=0.5pt,smooth]
	table[row sep=crcr]{
	134.4   0.99999567\\136.8   0.99996875\\139.2   0.9998337\\141.6   0.99931731\\144   0.99775329\\146.4   0.99387423\\148.8   0.98576568\\151.2   0.97111974\\153.6   0.94776372\\156   0.91426323\\158.4   0.87034762\\160.8   0.81700479\\163.2   0.75625342\\
	};
		\addlegendentry{\small CP, $\tilde{\rho}_{\rm si}\! = \!-\!27.5$}
		
		\addplot [awesome, line width=0.75pt,smooth,dashed]
table[row sep=crcr]{
134.4   0.99998042\\136.8   0.99988556\\139.2   0.99949291\\141.6   0.99822439\\144   0.99490965\\146.4   0.98768382\\148.8   0.97418828\\151.2   0.95209057\\153.6   0.91973417\\156   0.87663921\\158.4   0.82366096\\160.8   0.7627883\\
};
	\addlegendentry{\small CP, $\tilde{\rho}_{\rm si}\! = \!-\!12.5$}

		\addplot [orange, line width=1pt,dotted]
table[row sep=crcr]{
	9.6   1\\12   1\\14.4   1\\16.8   1\\19.2   1\\21.6   1\\24   1\\26.4   1\\28.8   1\\31.2   1\\33.6   1\\36   1\\38.4   1\\40.8   1\\43.2   1\\45.6   1\\48   1\\50.4   1\\52.8   1\\55.2   1\\57.6   1\\60   1\\62.4   1\\64.8   1\\67.2   1\\69.6   1\\72   1\\74.4   1\\76.8   1\\79.2   1\\81.6   1\\84   1\\86.4   1\\88.8   1\\91.2   1\\93.6   1\\96   1\\98.4   1\\100.8   0.99999991\\103.2   0.99999752\\105.6   0.99996509\\108   0.99970735\\110.4   0.99839164\\112.8   0.99372636\\115.2   0.98145894\\117.6   0.95619818\\120   0.91357652\\122.4   0.85239519\\124.8   0.77532208\\
};
		\addlegendentry{\small CP, $\tilde{\rho}_{\rm si}\! = \!2.5$}

		\addplot [darklavender, line width=0.75pt,dashdotted]
table[row sep=crcr]{
	9.6   1\\12   1\\14.4   1\\16.8   1\\19.2   1\\21.6   1\\24   1\\26.4   1\\28.8   1\\31.2   1\\33.6   1\\36   1\\38.4   1\\40.8   1\\43.2   1\\45.6   1\\48   1\\50.4   1\\52.8   1\\55.2   1\\57.6   1\\60   1\\62.4   1\\64.8   1\\67.2   1\\69.6   1\\72   1\\74.4   1\\76.8   1\\79.2   1\\81.6   1\\84   1\\86.4   1\\88.8   1\\91.2   1\\93.6   0.99999999\\96   0.9999998\\98.4   0.99999772\\100.8   0.99998151\\103.2   0.99988946\\105.6   0.99949473\\108   0.99817145\\110.4   0.99459419\\112.8   0.98657377\\115.2   0.97127924\\117.6   0.94590831\\120   0.90856057\\122.4   0.85890489\\124.8   0.79834792\\127.2   0.72968616\\
};
			\addlegendentry{\small ZP}

					\addplot [color=applegreen, line width=0.5pt, only marks, mark=o, mark options={solid},mark size=2.5]
table[row sep=crcr]{
139.2   0.9997418\\141.6   0.99906788\\144   0.99721184\\146.4   0.99291319\\148.8   0.98546\\151.2   0.97114\\153.6   0.94904\\156   0.91448\\158.4   0.86888\\160.8   0.81838\\163.2   0.75544\\
};
			\addlegendentry{\small CP, simulation}
			
			\addplot [color=darklavender, line width=0.5pt, only marks, mark=triangle, mark options={solid},mark size=2.5]
table[row sep=crcr]{
	9.6   1\\12   1\\14.4   1\\16.8   1\\19.2   1\\21.6   1\\24   1\\26.4   1\\28.8   1\\31.2   1\\33.6   1\\36   1\\38.4   1\\40.8   1\\43.2   1\\45.6   1\\48   1\\50.4   1\\52.8   1\\55.2   1\\57.6   1\\60   1\\62.4   1\\64.8   1\\67.2   1\\69.6   1\\72   1\\74.4   1\\76.8   1\\79.2   1\\81.6   1\\84   1\\86.4   1\\88.8   0.99999998\\91.2   0.99999983\\93.6   0.99999887\\96   0.99999366\\98.4   0.99996963\\100.8   0.99987556\\103.2   0.99956112\\105.6   0.99865755\\108   0.99640659\\110.4   0.99149895\\112.8   0.98738\\115.2   0.97098\\117.6   0.94656\\120   0.90998\\122.4   0.86102\\124.8   0.79722\\127.2   0.72654\\
};
			\addlegendentry{\small ZP, simulation}
		
					\addplot [color=applegreen, line width=0.5pt, only marks, mark=o, mark options={solid},mark size=2.5]
table[row sep=crcr]{
136.8   0.99980901\\139.2   0.99927031\\141.6   0.99771164\\144   0.99395067\\146.4   0.98844\\148.8   0.97456\\151.2   0.9512\\153.6   0.91996\\156   0.87688\\158.4   0.82294\\160.8   0.76534\\
};

\addplot [color=applegreen, line width=0.5pt, only marks, mark=o, mark options={solid},mark size=2.5]
table[row sep=crcr]{
103.2   0.99997465\\105.6   0.99981522\\108   0.99904728\\110.4   0.99633214\\112.8   0.99032\\115.2   0.97284\\117.6   0.94488\\120   0.9002\\122.4   0.8384\\124.8   0.76186\\
};
		
	\end{axis}

	%
	%

\end{tikzpicture}%
	\vspace{-0.7cm}
	\caption{PD comparison of CP and ZP systems over distance for the mmWave setting of Tab.~\ref{tab:range_comparison} and $N_{\rm zp}=128$ (or $N_{\rm oh} = 0.11$).}
	\label{fig:distance_ratio2}
\end{figure}
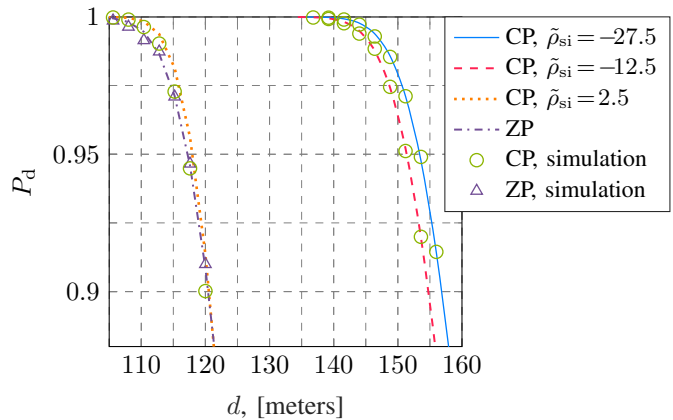

Another relevant analysis is to find the self-interference level $\rho_{\rm si}$ that makes $\delta(\rho_{\rm si})=1$, implying an equal range for the ZP and CP systems.
These values are found as $5.22$, $2.5$ and ${-1.14\, {\rm dB}}$ for $N_{\rm oh}$ being $5.9\%$, $11\%$ and $20\%$, respectively.
Again, considering the commonly value of as $N_{\rm oh}$ of $20\%$, even a very good FD cancellation with interference power ${-1.6\, {\rm dB}}$ in relation to the AWGN power leads to no range gain for the CP system.
In summary, the smaller the size of ZP in relation to the FFT size, the more imperfect the SI of the CP system can be. 

Lastly, complementary results of the PD curves over distance for the ZP and CP systems are plotted in Fig.~\ref{fig:distance_ratio2} to exemplify the above results.
For this analysis, we selected $N_{\rm zp} = 128$ (or $N_{\rm oh} = 0.11$) which corresponds to the middle line in Fig.~\ref{fig:distance_ratio}.
From Fig.~\ref{fig:distance_ratio2}, we observe that $d_{\rm zp} \approx \SI{120}{\m}$ at $P_{\rm d} = 0.9$.
For the CP system with $\tilde{\rho}_{\rm si} = -27.5 \, {\rm dB}$, we have $d_{\rm cp} \approx \SI{157}{\m}$ at $P_{\rm d} = 0.9$.
By doing ${d_{\rm cp}}/{d_{\rm zp}} \approx 1.3$, which matches the associated result of Fig.~\ref{fig:distance_ratio}.
Lastly, for $\tilde{\rho}_{\rm si} = 2.5 \, {\rm dB}$, both ZP and CP have the same range.
This result is in accordance to the results reported in Fig.~\ref{fig:distance_ratio}, where $\delta(\rho_{\rm si})=1$ for $\tilde{\rho}_{\rm si} = 2.5 \, {\rm dB}$.

\begin{table}[t!]
	\centering
	\caption{Parameters for results of Fig.~\ref{fig:pd_ub2}.}	\vspace{-0.3cm}
	\begin{tabular}{lccc}
		\toprule
		Param.		& sub-$\SI{6}{\GHz}$ & {mmWave} & {sub}-$\SI{}{\THz}$ \\ 
		\midrule
		frequency, $f_c$	& $\SI{2.4} \GHz$ & $\SI{24} \GHz$ & $\SI{140} \GHz$ \\
		bandwidth $B$	& $ \SI{100}\MHz$ 	&  $ \SI{1}\GHz$ &  $ \SI{4}\GHz$	\\
		antenna gain, $G$	& $16$ 			& $64$ & $128$	\\
		radar cross-section, $\sigma_{\rm RCS}$		&$\SI{10}\m^2$ & $\SI{10}\m^2$ &$\SI{1}\m^2$ \\
		transmit power, $P$	& $20 {\rm dBm}$  	& $20 {\rm dBm}$  & $20 {\rm dBm}$  \\
		FFT size, $N_{\rm f}$  & $1024$    & $1024$ & $1024$  \\
		\bottomrule
	\end{tabular}
	\label{tab:range_comparison}
	\vspace{-0.3cm}
\end{table}

\subsection{Clutter Analysis}\label{subsec:numerical_results_clutter}
\begin{figure}[t!]
	\centering
	\input{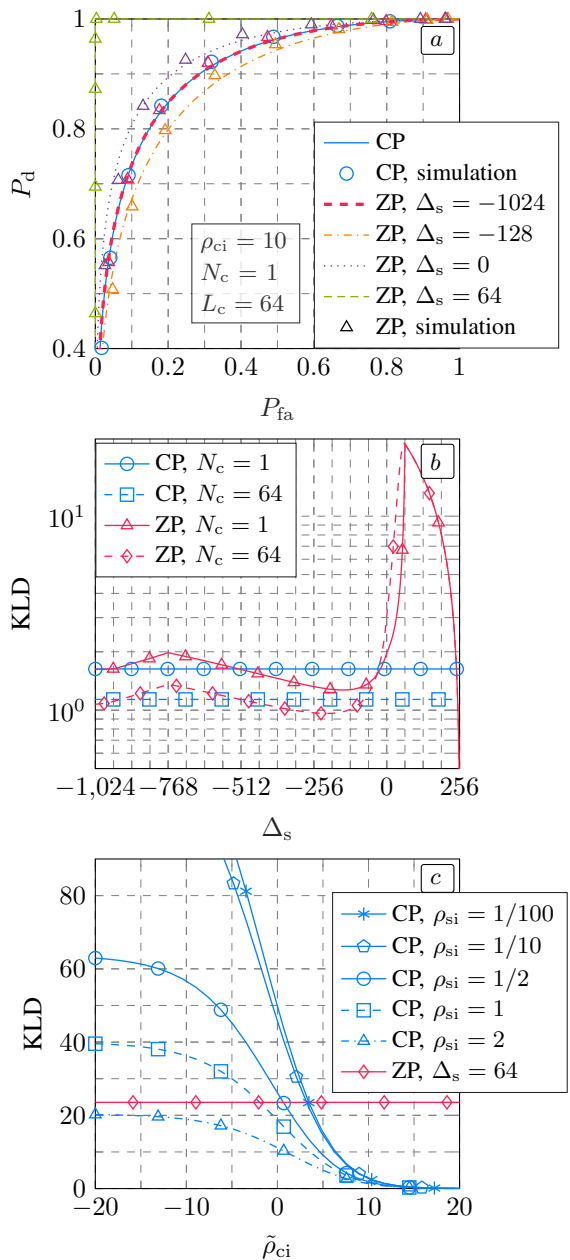}
	\caption{\textcolor{my_color}{Clutter analysis. The KLD is computed via \eqref{eq:kl_divergence_gaussian}. The theoretical PD, PFA considers the Gaussian approximations presented in Sections \ref{sec:statistics_zp} and \ref{sec:statistics_cp} for ZP and CP-OFDM, respectively.}}
	\label{fig:clutter_analysis}
	\vspace{-0.3cm}
\end{figure}
\textcolor{my_color}{In this subsection we analyze the capability of clutter rejection of the ZP-OFDM system for clutters that are closer than the target, as discussed in the Subsection \ref{subsec:clutter_rejection}.
	The parameters for this analysis are taken mmWave configuration of Table~\ref{tab:range_comparison} with antenna gain $G=10$.
	The FFT size and CP/ZP sizes are $N_{\rm f} = 1024$ and $N_{\rm cp}=N_{\rm zp}=256$, respectively.
	The target delay is set to $L_{\rm t} = 256$.}

\textcolor{my_color}{The results are depicted in the Figure \ref{fig:clutter_analysis}. 
	The top graph, $a$, analyzes the impact $\Delta_{\rm s}$  in terms of the \ac{ROC} curves, where the theoretical PFA and PD are based on the Gaussian approximations presented in Sections \ref{sec:statistics_zp} and \ref{sec:statistics_cp} for ZP and CP-OFDM, respectively. The simulation values are plotted for completeness.
	The scenario has $N_{\rm c} = 1$ with RCI power $\rho_{\rm ci} = 10$ and delay $L_{\rm c} = 64$.
	The RSI is $\rho_{\rm si}=1$ so that the interference is dominated by the clutter.
	We first observe that ZP-OFDM with $\Delta_{\rm s} = 0$ slightly outperforms CP-OFDM due to RSI of $\rho_{\rm si}=1$, but it is still far from the optimal ROC point $(P_{\rm fa}=0,P_{\rm d}=1)$.
	Changing $\Delta_{\rm s}$ to 64, ZP-OFDM also rejects the clutter, which improves its performance considerably reaching the optimal ROC point.
	In addition, the value of $\Delta_{\rm s} = -1024$ is also analyzed to study the performance in case ZP-OFDM has FD operation.
	In this case, ZP-OFDM performs equally to CP-OFDM. 
	This result indicates that when ZP and CP-OFDM are under the same interference levels, they perform equally.}

\textcolor{my_color}{The middle graph, $b$, compares the ZP and CP-OFDM in terms of the KLD of their approximated Gaussian PDFs under $H_0$ and $H_1$ as \eqref{eq:kl_divergence_gaussian}.
	In particular, the impact of $\Delta_{\rm s}$ is assessed for the same parameters of graph $a$, with the addition of $N_{\rm c} = 64$ clutters where the $q$th clutter has delay $L_{{\rm c}_q} = q$.
	The results show that the KLD of ZP-OFDM is maximized with $\Delta_{\rm s} = 64$.
	Interestingly, the KLD of ZP-OFDM does not vary monotonically with $\Delta_{\rm s}$.
	For instance, $\Delta=-128$ for ZP-OFDM has worse performance than the CP-OFDM, which can also be observed by the ROC curve in the graph $a$.
	This outcome confirms that a lower KLD implies in a lower ROC and vice-versa.
	When the number of clutter is increased\footnote{The total residual power of the clutters is the same since each clutter has power $\rho_{\rm ci}/N_{\rm c}$.}, the performance of CP-OFDM is decreased due to increased variance of the decision variable, which is caused by the additional covariance terms.
	Obviously, the ZP-OFDM system is not affected by this change for $\Delta_{\rm s}\geq 64$ clutter is rejected.}

\textcolor{my_color}{Lastly, the bottom graph, $c$, investigates the KLD of ZP/CP-OFDM for different values of RSI and RCI where $N_{\rm c} = 1$ and $L_{\rm c} = 64$.
	For the ZP-OFDM, $\Delta_{\rm s} = 64$ so that the KLD is invariant over $\tilde{\rho}_{\rm si} = 10\log_{10}(\rho_{\rm si})$.
	It shows that RCI has more impact on CP-OFDM when the RSI is small.
	For instance, the KLD is dropped considerably more for RSI $\rho_{\rm si}<1/2$ than $\rho_{\rm si}<2$.
	One useful analysis is to evaluate the value of $\tilde{\rho}_{{\rm ci}_{\rm min}}$ for which the KLD of CP-OFDM drops below the ZP-OFDM.
	In particular, for $\rho_{\rm si} = \left\{1,1/2,1/10\right\}$, $\tilde{\rho}_{{\rm ci}_{\rm min}}$ is found as $\left\{-1.16,1.16,2.52\right\} {\rm dB}$.}

\textcolor{my_color}{In summary, the above results have demonstrated the clutter rejection capability of ZP-OFDM.
	When CP-OFDM is employed, both RSI and RCI should be taken into account in order to determine if its performance loss is sufficiently high such that ZP-OFDM provides better detection performance.}

\section{Conclusion}\label{sec:conclusion}
In this paper, we have conducted a thorough study on the monostatic radar functionality of ISAC systems using ZP-OFDM focusing on the non-coherent energy detection method.
\textcolor{my_color}{Due to silent periods of ZP, the radar system can reject the self-interference due to full duplex operation, and nearby clutter.}

We have derived exact and approximate probability of detection (PD) curves for both ZP and CP systems, which have been used to generate new results.
In particular, an upper bound PD expression for the ZP system has been derived, which was used to evaluate the optimal ZP size that has been shown to be tightly connected to the system bandwidth.
Another main result has been to provide an expression to compare the detection range of the ZP and CP systems for a given PD. 
As an example, when the ZP size represents 25\% of the FFT size, the range achieved by the CP system is only 17\% larger than the ZP system if a perfect interference cancellation is implemented at the PD of 0.9.
The results also show that if the residual self-interference power level is $2.5\, {\rm dB}$ less than the AWGN power level, the CP and ZP systems have the same range at the PD of 0.9.
This indicates that in practice it may be advantageous to sacrifice range detection for simpler hardware implementation depending on the system parameters since the loss of the ZP system is not significant.
Lastly, we have shown the capability of ZP-OFDM to reject nearby clutter.

\appendix
%
		\subsection{Gamma Distribution}\label{apsubsec:gamma_statistics}
		Let $\mathbf{V} \sim \mathcal{CN}(\mathbf{0}_k,\mathbf{I}_k \theta)$ be a $k$ size vector of independent complex Gaussian \ac{RV} with zero mean and variance $\theta$.
		It is known that 
		\begin{equation}\label{eq:gamma_rv}
			\small
			G = \mathbf{V}^{\rm H}\mathbf{V} \sim \Gamma(k,\theta)
		\end{equation}
		follows the Gamma distribution with shape $k$ and scale $\theta$ as parameters, which is a generalization of the Chi-square distribution. Its \ac{CDF} is defined as
		\begin{equation}\label{eq:gamma_cdf}
			\small
			F_G(\lambda|k,\theta) = \frac{\gamma(k,\lambda/\theta)}{\Gamma(k)},
		\end{equation}
		for $\lambda \geq 0$, where $\gamma(\cdot,\cdot)$ and $\Gamma(\cdot)$ are the incomplete Gamma and Gamma functions. 
		The moments of $G$ are given by
		\begin{equation}\label{eq:gamma_moments}
			\small
			\mathbb{E}(G) = k\theta \,\,\,\,\,\,  \text{and} \,\,\,\,\,\,  \mathbb{V}(G) = k\theta^2.
		\end{equation}	
		The inverse of \eqref{eq:gamma_cdf} is also of interest and is defined as
		\begin{equation}\label{eq:gamma_icdf}
			\small
			\textcolor{my_color}{\begin{split}
					\lambda  \!= \!\{\lambda : F_G(\lambda|k,\theta) = x \} \!= \!F_G^{-1}(x|k,\theta) \!= \!F_G^{-1}(x|k,1)\theta .
			\end{split}}
		\end{equation}
		Although an analytical solution for $F_G^{-1}(x|k,\theta)$ is not known, an iterative approach based on Newton's method that converges to the solution is used.
		\textcolor{my_color}{Finally, we highlight the {\it scaled threshold property} of \eqref{eq:gamma_icdf} that implies which writes $F_G^{-1}(x|k,1)\theta$
			due to the fact that $\lambda$ is scaled by $\theta^{-1}$ in \eqref{eq:gamma_cdf}.}
		\subsection{Sum of two i.i.d. Gamma RVs}\label{apsubsec:sum_gamma} 
		Let $G_1\sim \Gamma(k_1,\theta_1)$ and $G_2\sim\Gamma(k_2,\theta_2)$ be two independent Gamma RVs. 
		Consider the sum $S=G_1 + G_2$, the \ac{PDF} of $S$ is given as \cite{Bausch_2013}
		\begin{equation}\label{eq:sum_gamma_pdf}
			\small
			\begin{split}
				f_{S}(s|&k_1,\theta_1,k_2,\theta_2) = \\  &\frac{s^{k_1+k_2-1}e^{-s/\theta_2}}{\theta_1^{k_1}\theta_2^{k_2}\Gamma(k_1+k_2)}{}_1\text{F}_1\left(k_1;k_1\!+\!k_2;(\theta_2^{-1}\!-\!\theta_1^{-1})s\right)
			\end{split}
		\end{equation}
		for $s\geq 0$, where ${}_1 \text{F}_1(\cdot;\cdot;\cdot)$ is a confluent hypergeometric function (Kummers function of the first kind).
		To our knowledge, there is no analytical solution to the integral \eqref{eq:sum_gamma_pdf}, we compute the CDF
		\begin{equation}\label{eq:sum_gamma_cdf}
			\small
			F_S(\lambda|k_1,\theta_1,k_2,\theta_2) = \int_0^{\lambda} f_{S}(s|k_1,\theta_1,k_2,\theta_2) ds
		\end{equation}
		numerically.
		\begin{figure*}[b]
			\hrulefill
			\begin{equation}\label{eq:C_zp_j}
				\small
				\textcolor{my_color}{	C_{{\rm zp}_j} = \left\{\begin{matrix}
						-\Delta_s, & \Delta_s < \dot{L}_j - N_{\rm zp}- N_{\rm f} \\
						\max(\min(-\Delta_s+\min(\dot{L}_j,N_{\rm zp}), \min(-\dot{L}_j+N_{\rm zp},0)+N_{\rm f}),0) , & {\rm otherwise} 
					\end{matrix}\right.}
			\end{equation}
			\begin{equation}\label{eq:Cjj}
				\small
					\textcolor{my_color}{	C_{{\rm zp}_{j,j'}} =  \min(\max ( N_{\rm zp} + \dot{L}_j \!+ \!\min (-\Delta_{\rm s},N_{\rm f} \!- \!\dot{L}_{j'},0),0 ), \max (\dot{L}_j, \min(\Delta_{\rm s},N_{\rm f}\!-\!\dot{L}_j),0))}
				\end{equation}
			\end{figure*}
			\textcolor{my_color}{\subsection{Moments of $Z_{\rm zp}$}\label{apsubsec:moments_ZP}
				\subsubsection{Mean}
				The expected value of $z_n$ in \eqref{eq:zn_zp}  is computed as
				\begin{equation}\label{eq:mean_zn}
					\small
					\mathbb{E}(z_{{\rm zp}_{n}}) = \sum_{j=0}^{N_{\rm c} + 1} \eta \dot{h}_j^2 I_{j,n}  + \sigma^2
				\end{equation}
				using the fact that $\mathbb{E}(\mathbf{x}_{\rm zp}[n] \mathbf{x}_{\rm zp}[m]^\dagger) = 0$ for $n \neq m$.
				The indicator variable $I_{j,n}$ is
				\begin{equation}\label{eq:Ijn}
					\small
					I_{j,n} = \left\{\begin{matrix}
						0, & \langle n-\dot{L}_j \rangle_N > N_{\rm f} \\  
						1, & {\rm otherwise}
					\end{matrix}\right.,
				\end{equation}
				which basically selects the channel $\dot{h}_j^2$ if the ZP signal does not contain zero in the $(n-\dot{L}_j)$th sample where $\dot{L}_j$ is simply the corresponding delay to the channel $\dot{h}_j^2$ in \eqref{eq:h_dot}.
				Lastly, we note that the summation term in \eqref{eq:mu_zp} allows the following simplification 
				\begin{equation}\label{eq:mu_zp_simplification}
					\frac{1}{N_{\rm zp}-\Delta_{\rm s}}\sum_{n = N_{\rm f} + \Delta_{\rm s} + 1}^{N-1}\mathbb{E}(z_{{\rm zp}_{n}}) =  \sum_{j = 0}^{N_{\rm c}+1}\dot{h}_j^2 C_{{\rm zp}_j} + \sigma^2,
				\end{equation}
				where $C_{{\rm zp}_j} \in \mathbb{N}_0$ is given in \eqref{eq:C_zp_j} which counts the number of occurrences where the indicator variable $I_{j,n}$ is non zero in \eqref{eq:mean_zn}.
				The approach to finding $C_{{\rm zp}_j}$ was plotting it for different configurations and then matching a corresponding equation to its graph.
				The details of this process are omitted due to the lack of space. 
				The functions $\max(\cdot)$ and $\min(\cdot)$ are needed to set inferior and superior limits to respect the ZP window. 
				\subsubsection{Variance}
				Clearly, $z_{{\rm zp}_{n}}$ in \eqref{eq:zn_zp} is a Gamma RV as \eqref{eq:gamma_rv} with shape $k=1$, whose variance is given by 
				\begin{equation}
					\small
					\mathbb{V}(z_{{\rm zp}_{n}}) = \mathbb{E}(z_{{\rm zp}_{n}})^2
				\end{equation}
				according to \eqref{eq:gamma_moments}, and $\mathbb{E}(z_{{\rm zp}_{n}})$ is given by \eqref{eq:mean_zn}.
				\subsubsection{Covariance}
				Using the fact that $\mathbb{E}(\mathbf{x}_{\rm zp}[n] \mathbf{w}[m]^\dagger) = 0$ for all $(n,m)$ pairs, the general covariance between $z_{{\rm zp}_{n}}$ and $z_{{\rm zp}_{m}}$ is found as
				\begin{equation}
					\small
					\mathbb{C}(z_{{\rm zp}_{n}},z_{{\rm zp}_{m}}) =  \sum_{j=0}^{N_{\rm c} + 1} \sum_{j' = j+1}^{N_{\rm c} + 1} \eta^2 \dot{h}_j^2 \dot{h}_{j'}^2 I_{j,j',n,m},
				\end{equation}
				where the indicator function $I_{j,j',n,m}$ is given by
				\begin{equation}\label{eq:I_jjnm}
					\small
					I_{j,j',n,m} = \left\{\begin{matrix}
						&  					&  n-\dot{L}_j = m-\dot{L}_{j' }					 \\
						1,  &   {\rm and} &   \langle n-\dot{L}_j \rangle_N > N_{\rm f}   \\
						&   {\rm and} &   \langle m-\dot{L}_{j'} \rangle_N > N_{\rm f}      \\
						0,  & & {\rm otherwise}
					\end{matrix}\right..
				\end{equation}
				for $0 \leq n < m \leq N_{\rm f}-1$ and selects the pairs of channels $(j,j')$ when i) the samples $n-\dot{L}_j$ and $m-\dot{L}_{j'}$ coincide and ii) they correspond to non zero samples of the ZP signal.
				Assuming that the clutter delays $L_{{\rm c}_{m}} < N_{\rm zp}$, the contribution of the covariance terms in \eqref{eq:var_zp} can be further simplified as
				\begin{equation}\label{eq:cov_simplification}
					\small
					\begin{split}
						\sum_{n = N_{\rm f} + \Delta_{\rm s}}^{N-1}  \sum_{m=n+1}^{N-1}\mathbb{C}(z_{{\rm zp}_{n}},z_{{\rm zp}_{m}}) = \sum_{j=0}^{N_{\rm c}+1}\sum_{j'=j+1}^{N_{\rm c}+1}C_{{\rm zp}_{j,j'}} \eta^2 \dot{h}_j^2 \dot{h}_{j'}^2,
					\end{split}
				\end{equation}
				where the quantity $C_{{\rm zp}_{j,j'}} \in \mathbb{N}_0$ is given in \eqref{eq:Cjj} and counts the amount of times in which $I_{j,j',n,m}$ equals one for a given pair $(j,j')$.
				A detailed demonstration of \eqref{eq:Cjj} is omitted due to the lack of space.
				Similarly to \eqref{eq:C_zp_j}, the functions $\max(\cdot)$ and $\min(\cdot)$ are needed to set inferior and superior limits to respect the ZP window.}
			\subsection{Computing SNR Expressions Depending on PD}\label{sec:ap_distance_zp}
			\subsubsection{ZP-OFDM} 
			We can relate the SNR to the PD due to the fact that the scale parameter $\theta_{\rm t} = ( |h_{\rm r}|^2\eta+\sigma^2)/N_{\rm zp}$ in \eqref{eq:E_r_param} can be related to the SNR and $P_{\rm d}$ as
			\begin{equation}\label{eq:theta_r}
				\small
				\begin{split}
					\theta_{\rm t}  \stackrel{(a)}{=} (\rho_{{\rm zp}}+1)\frac{\sigma^2}{N_{\rm zp}}
					\stackrel{(b)}{\approx} \frac{\lambda_{\rm zp}}{F^{-1}_{Z_{\rm zp}|H_1}(1-P_{\rm d}|k_{\rm t},1)}
				\end{split}
			\end{equation}
			where equality $(a)$ is obtained by using the SNR $\rho_{{\rm zp}}$ in \eqref{eq:rho_zp}.
			The approximation $(b)$ is obtained by combining the approximate PD in \eqref{eq:pd_zp_gamma_approx} with \eqref{eq:gamma_icdf}.
			\textcolor{my_color}{Then, combining $(a)$ and $(b)$ in \eqref{eq:theta_r} leads to
			\begin{equation}\label{eq:app_pd_zp}
				\small
				\begin{split}
					\rho_{{\rm zp}}  \approx \frac{N_{\rm zp}\lambda_{\rm zp}}{\sigma^2 F^{-1}_{Z_{\rm zp}|H_1}(1-P_{\rm d}|N_{\rm zp},1)} - 1,
				\end{split}
			\end{equation}	
			where $k_{\rm t}$ has been replaced by $N_{\rm zp}$.
			It is worth recalling that for $\tilde{L}=N_{\rm zp}$, the Gamma distribution is exact according to \eqref{eq:pd_zp_exact}. 
			Now, we show that \eqref{eq:app_pd_zp} depends only on $N_{\rm zp}$ for a constant PFA by replacing 
				\begin{equation}\label{eq:lambda_app}
					\small
					\begin{split}
						\lambda_{\rm zp}   = F^{-1}_{Z_{\rm zp}|H_0}(1-P_{\rm fa}|N_{\rm zp},1)\sigma^2/N_{\rm zp}  = \lambda_{\rm norm_{\rm zp}}\sigma^2/N_{\rm zp},
					\end{split}
				\end{equation}
				by setting $k=N_{\rm zp}$ and $\theta_{\rm w} = \sigma^2/N_{\rm zp}$ as \eqref{eq:E_w_param} since $h_{\rm t}=0$ under $H_0$, where \eqref{eq:gamma_icdf} has been used.
				In \eqref{eq:lambda_app}, $\lambda_{\rm norm_{\rm zp}} = F^{-1}_{Z_{\rm zp}|H_0}(1-P_{\rm fa}|N_{\rm zp},1)$ is the threshold that achieves a given $P_{\rm fa}$ for $\theta=1$.
				Basically, \eqref{eq:app_pd_zp} can be interpreted as the necessary SNR to achieve a given $P_{\rm fa}$ and $P_{\rm d}$ pair.}
			\subsubsection{CP-OFDM} 
			\textcolor{my_color}{The PD for CP-OFDM can also be approximated to the Gamma distribution as \eqref{eq:Pd_cp_approx}.
				For the general case $|h_{\rm i}|^2 > 0$, the scale parameter can be written as $\theta_{\rm t} = (\rho_{{\rm zp}}+1)(1+\rho_{\rm si})\sigma^2/N$ that has $N$ instead of $N_{\rm zp}$ and the multiplicative factor $(1+\rho_{\rm si})$. 
				Then, following the same steps as the ZP case, its analogous SINR expression is given by}
			\begin{equation}
				\small
				\textcolor{my_color}{\rho_{{\rm cp}} \approx \frac{\lambda_{\rm norm_{\rm cp}}}{F^{-1}_{Z_{\rm cp}|H_1}(1-P_{\rm d}|N,1)} - 1.}
			\end{equation}	

\bibliography{references_ha}{}
\bibliographystyle{ieeetr}

\vfill

\end{document}